# TITANITE-BEARING CALC-SILICATE ROCKS CONSTRAIN TIMING, DURATION AND MAGNITUDE OF METAMORPHIC $CO_2$ DEGASSING IN THE HIMALAYAN BELT

Giulia Rapa[a], Chiara Groppo[a,b,*], Franco Rolfo[a,b], Maurizio Petrelli[c], Pietro Mosca[b], Diego Perugini[c]

[a]Department of Earth Sciences, University of Torino, Via Valperga Caluso 35, 10125 Torino, Italy

[b]IGG-CNR, Via Valperga Caluso 35, 10125 Torino, Italy

[c]Department of Physics and Geology, University of Perugia, Piazza Università, 06100 Perugia, Italy

*Corresponding Author

Chiara Groppo

Dept. of Earth Sciences, University of Torino

Via Valperga Caluso, 35 – 10125 Torino, Italy

Tel. +39 0116705106

Fax +39 0116705128

**Abstract**

The pressure, temperature, and timing (P-T-t) conditions at which $CO_2$ was produced during the Himalayan prograde metamorphism have been constrained, focusing on the most abundant calc-silicate rock type in the Himalaya. A detailed petrological modelling of a clinopyroxene + scapolite + K-feldspar + plagioclase + quartz ± calcite calc-silicate rock allowed the identification and full characterization - for the first time – of different metamorphic reactions leading to the simultaneous growth of titanite and $CO_2$ production. The results of thermometric determinations (Zr-in-Ttn thermometry) and U-Pb geochronological analyses suggest that, in the studied lithology, most titanite grains grew during two nearly consecutive episodes of titanite formation: a near-peak event at 730-740°C, 10 kbar, 30-26 Ma, and a peak event at 740-765°C, 10.5 kbar, 25-20 Ma. Both episodes of titanite growth are correlated with specific $CO_2$-producing reactions and constrain the timing, duration and P-T conditions of the main $CO_2$-producing events, as well as the amounts of $CO_2$ produced (1.4-1.8 wt% of $CO_2$). A first-order extrapolation of such $CO_2$ amounts to the orogen scale provides metamorphic $CO_2$ fluxes ranging between 1.4 and 19.4 Mt/yr; these values are of the same order of magnitude as the present-day $CO_2$ fluxes degassed from spring waters located along the Main Central Thrust. We suggest that these metamorphic $CO_2$ fluxes should be considered in any future attempts of estimating the global budget of non-volcanic carbon fluxes from the lithosphere.

## 1. Introduction

Metamorphic degassing from active collisional orogens supplies a significant fraction of $CO_2$ to the atmosphere, playing a key role in the long-term (> 1 Ma) global carbon cycle (e.g. Bickle, 1996; Evans, 2011; Gaillardet and Galy, 2008; Skelton, 2011). Primary geologic settings for the production of significant amounts of metamorphic $CO_2$ include "large-hot" collisional orogens, where decarbonation reactions occur at a relatively high temperature within carbonate-bearing metasediments (e.g. calc-silicate rocks), in which metamorphic reactions between carbonates and silicates trigger $CO_2$ production.

Being the most prominent recent and still active "large-hot" orogen on Earth, the Himalayan belt is the best candidate for the generation of a significant amount of metamorphic $CO_2$ during the Cenozoic. Previous studies aimed at quantifying metamorphic $CO_2$ degassing in the Himalayas yield conflicting results and argued either in favour of a significant contribution to periods of greenhouse global warming in the past (Becker et al., 2008; Kerrick and Caldeira, 1993), or against it (Kerrick and Caldeira, 1998, 1999; Selverstone and Gutzler, 1993). The nature and magnitude of the metamorphic $CO_2$ cycle in the Himalaya, however, is still poorly known because, as pointed out by Kohn and Corrie (2011), Himalayan calc-silicate rocks have been largely ignored in previous metamorphic studies. Major uncertainties are: (1) the volumes of $CO_2$-source rocks and their relevant mineral assemblages; (2) the nature of the metamorphic $CO_2$-producing reactions; (3) the P-T conditions at which the decarbonation reactions took place; (4) the timing and duration of $CO_2$-producing events, which is essential for understanding whether the Himalayan orogeny was responsible for ancient periods of global warmth and for estimating the annual mass flux of $CO_2$ released during these periods.

Concerning point (1), although calc-silicate rocks in the metamorphic core of the Himalaya (i.e. in the Greater Himalayan Sequence: GHS) have been reported since the pioneering works of Bordet (1961), Gansser (1964) and Le Fort (1975), a systematic study of the different types of calc-silicate rocks and of their distribution within the GHS has been undertaken only recently (Rolfo et al., 2015, 2017). Point (2) has also been a major focus of recent studies (Groppo et al., 2013, 2017): devolatilization reactions in different types of Himalayan calc-silicate rocks have been investigated using a novel approach which combines phase diagram modelling with a detailed interpretation of microstructures. New $CO_2$-producing reactions have been recognised, involving Ca-Mg-Fe- (garnet, clinopyroxene) and Ca-Na- (plagioclase, scapolite) solid solutions, as well as K-bearing phases (biotite, muscovite, K-feldspar). Groppo et al. (2017) also provided preliminary insights on point (4) by demonstrating that, in internally buffered systems, $CO_2$ is mainly produced at isobaric invariant points, thus suggesting that the duration of the $CO_2$-producing events is significantly shorter than the duration of prograde metamorphism.

This work focuses on points (3) and (4), i.e. on the still vague estimate of both the P-T conditions and the timing of $CO_2$-producing events which occurred during Himalayan prograde metamorphism. The P-T conditions experienced by calc-silicate rocks can be indirectly estimated using associated metapelites (e.g. Groppo et al., 2013, 2017). As a drawback, indirect P-T estimations provide only qualitative knowledge about the timing of the events. Alternatively, the recently calibrated trace element thermometer based on the zirconium content in titanite (Zr-in-Ttn: Hayden et al., 2008) is a promising method to directly constrain the thermal history of calc-silicate rocks, because titanite is a common accessory mineral in these lithologies and the Zr-in-Ttn thermometer is independent of $X(CO_2)$ in the fluid. Titanite's slow Zr diffusivity permits retention of Zr-temperatures up to at least 750-775 °C (Cherniak, 2006; Kohn and Corrie, 2011; Walters and Kohn, 2017; see also Kohn, 2016, 2017 for a review). The Zr-in-Ttn thermometer is, therefore, applicable to calc-silicate rocks from the GHS, which experienced peak temperatures $\leq$ 800°C (e.g. Groppo et al., 2013, 2017; Kohn and Corrie, 2011). A further advantage of titanite is that it can also be used as a chronometer (e.g. Frost et al., 2000; Kohn, 2017), thus linking U-Pb ages with temperatures through simultaneous geochronology and trace element thermometry in single microanalytical spots. Recent petrochronological studies have demonstrated that Pb diffusivity in titanite is very slow (e.g. Kohn and Corrie, 2011; Gao et al., 2012; Spencer et al., 2013; Stearns et al., 2015; Walters and Kohn, 2017). As a consequence, titanite crystals act as recorders of their prograde growth at high temperature. Few studies have already successfully obtained titanite U-Pb ages from Himalayan calc-silicate rocks and other Himalayan lithologies (e.g. Cottle et al., 2011; Kohn and Corrie, 2011; Warren et al., 2012; Walters and Kohn, 2017). However, although well constrained in term of T and P, the titanite ages obtained so far are still not related to specific $CO_2$-producing events.

The aim of this work is to constrain the P-T-t conditions at which $CO_2$ was produced during the Himalayan orogeny, focusing on the most abundant calc-silicate type in the Himalayas (i.e. clinopyroxene + scapolite + K-feldspar + plagioclase ± calcite calc-silicate rocks; Rolfo et al., 2017). Metamorphic reactions that led to the simultaneous growth of titanite and production of $CO_2$ are identified and fully characterized for the first time. T-t data from different titanite generations are correlated with specific $CO_2$-producing reactions, allowing constraining the timing, duration and P-T conditions of the main $CO_2$-producing events, as well as the amounts of $CO_2$ produced; furthermore, a first order extrapolation of these $CO_2$ amounts to the orogen scale is proposed.

## 2. Sample description

### *2.1 Geological setting*

The studied sample 14-53c was collected from a calc-silicate level (~ 2 m -thick) embedded within metapelitic gneisses of the Lower Greater Himalayan Sequence (L-GHS), in the Gosainkund-Helambu region of central Nepal (N28°00'37.3'', E85°29'50.9''; 3365 m a.s.l.; Fig. 1a,b). In this area, the L-GHS is bounded at its bottom by the Main Central Thrust (MCT), which juxtaposes the L-GHS onto the Lesser Himalayan Sequence (LHS), and at its top by the Langtang Thrust (LT: Kohn, 2008; Kohn et al., 2005), which juxtaposes the Upper GHS unit (U-GHS) onto the L-GHS. Both the L-GHS and the U-GHS are thick metasedimentary sequences consisting of medium- to high-grade metapelites with calc-silicate layers and bodies of granitic orthogneisses. Metamorphic grade increases upward, reaching anatexis in the structurally higher levels of the L-GHS and in the U-GHS. The structural and lithological features of the area are described in detail by Rapa et al. (2016). The metapelitic gneisses hosting the studied sample experienced peak P-T conditions of 720-760 °C, 9-10 kbar (e.g. sample 14-52: Rapa et al., 2016).

Sample 14-53c was chosen because it is representative of the most abundant calc-silicate type in the GHS, which occurs as tens to hundreds of meter thick layers in the U-GHS and as thinner (few meters to tens of meters thick) layers in the L-GHS (Rolfo et al., 2015, 2017). These calc-silicate gneisses belong to the complex group of metacarbonate rocks, ranging in composition from almost pure marble to carbonate-free lithologies (Fig. SM1). They derive from nearly isochemical regional metamorphism of former marly protoliths (Fettes and Desmons, 2007), whose main components were clay minerals (smectite, illite, kaolinite and/or chlorite), carbonates (mainly calcite) and quartz. Depending on the primary amount of carbonates, decarbonation reactions which occurred during prograde metamorphism can have led to the partial or total consumption of the original carbonate minerals, as observed in the calcite-poor domains of the studied sample (see Section 2.2). The sample was collected far from structures that can potentially represent preferential pathways for the entrance of external fluids (e.g. fold hinges, shear zones, faults, veins).

### 2.2 Petrography and mineral chemistry

The calc-silicate level from which the studied sample was collected shows a banded structure, being characterized by dm-thick layers with different mineral assemblages (Fig. 1c). These layers consist of: quartz + clinopyroxene + calcic plagioclase (sample 14-53a); garnet + calcic plagioclase + clinopyroxene + epidote (14-53b); quartz + clinopyroxene + calcic plagioclase + K-feldspar + scapolite ± calcite (14-53c); quartz + garnet + scapolite + clinopyroxene + calcic plagioclase (14-53d). The studied sample 14-53c is a fine-grained calc-silicate rock with a granoblastic texture, in which calcite-rich domains alternate with calcite-poor domains, defining a cm-thick layering (Fig. 2). The main mineral assemblage in both domains consists of quartz + clinopyroxene + calcic plagioclase + K-feldspar + scapolite ± calcite, minor epidote, biotite and titanite, and accessory bluish tourmaline and zircon; the modal proportion of the mineral phases is different in the two domains (Table 1a and Fig. 2). Representative microstructures and mineral chemical data are reported in Fig. 3-6. Mineral abbreviations are according to Whitney and Evans (2010).

K-feldspar ($Or_{95-100}$) in both domains includes quartz, rare scapolite and clinopyroxene (Fig. 3b-c), and is locally included in plagioclase (Fig. 4b,i); myrmeckitic microstructures are observed at the interface with plagioclase. In the Cal-rich layers, K-feldspar is locally included in biotite (Fig. 3a), showing corroded margins against it. Triple junctions between K-feldspar and scapolite are observed in the matrix. Clinopyroxene ($Di_{64-67}$) is pale green and occurs as mm-sized granoblasts. The clinopyroxene core contains rare small mono- and polymineralic inclusions of amphibole (mostly actinolite with $X_{Mg}$=0.65-0.90) + calcite + quartz + titanite + biotite with corroded margins, while the rim includes mono-mineralic inclusions of quartz, scapolite, titanite and K-feldspar (Fig. 3d-f). Clinopyroxene is rarely included in plagioclase in both domains (Fig. 4b,c).

Plagioclase ($An_{90-97}$) mainly occurs as medium-grained granoblasts. In the Cal-poor domains, plagioclase often includes vermicular quartz and mono-mineralic inclusions of scapolite, K-feldspar, and rare epidote (Fig. 4a,i). In the Cal-rich layers it includes rounded calcite, scapolite, clinopyroxene, K-feldspar, and quartz (Fig. 4b-d). A less calcic plagioclase ($An_{73-78}$) is rarely preserved as relict inclusions in scapolite and titanite (Fig. 4e & 5b). Plagioclase is locally surrounded by a rim of epidote in the Cal-poor domains (Fig. 4i), and of epidote + calcite in the Cal-rich domains. Scapolite in the matrix ranges from $eqAn_{61-68}$ in the Cal-poor domains, to $eqAn_{45-68}$ in the Cal-rich domains and contains some Cl (0.31-1.55 wt%). Matrix scapolite includes quartz, rare relict plagioclase ($An_{73-78}$) (Fig. 4e) and titanite, and is locally partially replaced by symplectitic aggregates of epidote + quartz ± calcite (Fig. 4g,h). Scapolite included in K-feldspar, plagioclase, and clinopyroxene, has a composition in the range $eqAn_{64-75}$ (equivalent anorthite content, eqAn = (Al-3)/3); a few inclusions in titanite (Fig. 5f) show lower eqAn values ($eqAn_{48}$).

Biotite ($X_{Mg}$=0.57-0.66; Ti=0.04-0.06 a.p.f.u.) in the matrix occurs as deformed lamellae with corroded margins; it is locally included in clinopyroxene and titanite (Fig. 3f & 5d-f), with lobate and rounded margins against the hosting mineral. Epidote included in plagioclase is a $Zo_{65}$. Epidote forming the symplectitic aggregates replacing scapolite is locally zoned, with a $Zo_{57-63}$ core and a $Zo_{81}$ rim. Epidote replacing plagioclase and clinopyroxene is a $Zo_{61-79}$.

Calcite is almost pure, with small amounts of Fe (0.01-0.02 a.p.f.u.) and Mg (0.01 a.p.f.u.). In both Cal-poor and Cal-rich domains, calcite is preserved as inclusion within clinopyroxene (in association with amphibole and quartz, Fig. 3d) and titanite (Fig. 5c,d); in the Cal-rich domains, it is also included in plagioclase (Fig. 4d). Calcite granoblasts in the matrix occur only in Cal-rich domains, where they locally overgrow scapolite (Fig. 4f).

## 3. Methods

### *3.1 Whole-rock microstructural and mineral chemical characterization*

Qualitative major element X–ray maps of the entire thin section were acquired using a micro-XRF Eagle III–XPL spectrometer equipped with an EDS Si(Li) detector and with an EdaxVision32 microanalytical system at the Department of Earth Sciences, University of Torino. The operating conditions were 100 ms counting time, 40 kV accelerating voltage and a probe current of 900 μA. A spatial resolution of about 65 μm in both x and y directions was used. Quantitative modal amounts of each mineral phase were obtained by processing the μ-XRF maps with the software program Petromod (Cossio et al., 2002).

The rock-forming minerals were analysed with a Jeol JSM-IT300LV Scanning Electron Microscope equipped with an energy dispersive spectrometry (EDS) Energy 200 system and an SDD X-Act3 detector (Oxford Inca Energy) at the Department of Earth Sciences, University of Torino. The operating conditions were 50 s counting time and 15 kV accelerating voltage. SEM-EDS quantitative data (spot size 2 μm) were acquired and processed using the Microanalysis Suite Issue 12, INCA Suite version 4.01; natural mineral standards were used to calibrate the raw data; a ϕρZ correction was applied.

The bulk rock compositions of the Cal-poor and Cal-rich domains were calculated by combining the mineral proportions obtained from the quantitative modal estimate of the micro-XRF maps (see above) with mineral chemistry acquired with the SEM–EDS and are given in Table 2. The fine-grained nature of epidote hampered its discrimination in the micro-XRF map. Therefore, its (low) modal amount was estimated visually.

### *3.2 Titanite textural, chemical and isotopic characterization*

The textural relationships between titanite and the other minerals, as well as the major element composition of titanite, were investigated by SEM (same equipment used for analysing the main rock-forming minerals; see section 3.1). Titanite is abundant in both Cal-poor and Cal-rich layers; it often includes scapolite, quartz and biotite, and less frequently plagioclase ($An_{73-94}$), K-feldspar, calcite, epidote and apatite (Fig. 5a-f). Titanite crystals are commonly 100 to 250 μm in length and 20 to 100 μm in width. Most of the analyzed titanite grains are dispersed in the rock matrix; a few analyses have been performed on titanite grains included in clinopyroxene and scapolite. Titanite shows a patchy zoning, as already observed by other authors (e.g. Franz and Spear, 1985; Hayden et al., 2008; Kohn and Corrie, 2011; Kohn, 2017).

In situ trace element analyses and U/Pb age determinations of titanite samples were performed by Laser Ablation Inductively Coupled Plasma Mass Spectrometry (LA-ICP-MS) at the Department of Physics and Geology, University of Perugia (Petrelli et al., 2016). The LA-ICP-MS system was a G2 Teledyne Photon Machine ArF excimer (193 nm) LA system coupled with an iCAPQ Thermo Fisher Scientific, quadrupole based, ICP-MS. LA-ICP-MS operating conditions were tuned before each analytical session to provide maximum signal intensity and stability for the ions of interest. Oxides formation (ThO/Th < 0.5) and Th/U sensitivity ratio (Th/U ~ 1) were also monitored to maintain robust plasma conditions (e.g. Petrelli et al., 2016). Because of the reduced dimensions of the analysed titanite samples, a laser beam diameter ranging between 30 and 40 μm was utilized. The laser energy on the sample surface and the repetition rate were fixed to 3.5 $J/cm^2$ and 8 Hz, respectively. In detail, trace elements determinations, performed on the isotopes $^{90}Zr$, $^{232}Th$ and $^{238}U$, were calibrated using the NIST SRM 612 glass reference material (Pearce et al., 1997). The OLT1 (Kennedy et al., 2010) and MKED1 (Spandler et al., 2016) titanite reference materials were employed as quality controls for trace element determinations. Data reduction was performed using the IOLITE 3 (Paton et al., 2011) software utilizing the protocol described by Longerich et al. (1996). U/Pb age determinations were performed using the same protocol reported by Chew et al. (2016). The OLT1 (Kennedy et al., 2010) and MKED1 (Spandler et al., 2016) titanite reference materials were employed as calibrator and quality control, respectively. Both U/Pb age determinations and trace element analyses were performed preferably in the most homogeneous portions of each titanite grain, as determined using BSE images. Results on unknown samples are reported in Table 3; results from the quality controls are reported in Tables SM1-SM2 and Fig. SM5.

### *3.3 Phase diagrams computation*

Phase diagrams in the $Na_2O$-CaO-$K_2O$-(FeO)-MgO-$Al_2O_3$-$SiO_2$-$TiO_2$-$H_2O$-$CO_2$ (NCK(F)MAST–HC) system were calculated using Perple_X (version 6.7.4, November 2016) (Connolly 1990, 2009) and the internally consistent thermodynamic dataset and equation of state for $H_2O$ of Holland and Powell (1998, revised 2004). $Fe^{3+}$ was neglected because $Fe^{3+}$-rich oxides are absent and the amount of $Fe^{3+}$ in the analysed minerals is very low. All the phase diagrams were calculated along a P/T gradient reflecting the P-T path followed by the hosting metapelites (Rapa et al., 2016).

For the calculation of the P/T-X($CO_2$) pseudosections in the NCKFMAST-HC system the following solid solution models were used: Holland and Powell (1998) for dolomite, chlorite, garnet and clinopyroxene, Diener and Powell (2012) for amphibole, Tajcmanova et al. (2009) for biotite, Newton et al. (1980) for plagioclase and Kuhn et al. (2005) for scapolite, in addition to the binary $H_2O$-$CO_2$ fluid. Calcite, K-feldspar, muscovite, paragonite, margarite, kyanite, quartz, titanite, rutile, ilmenite and zoisite were considered as pure end-members. In the pseudosection modelling, the binary $H_2O$-$CO_2$ fluid is considered as a saturated fluid phase.

P/T-X($CO_2$) grids and P-T mixed-volatile projection were calculated in the NCKMAST-HC system considering the following mineral phases: plagioclase, scapolite and biotite (phlogopite-Ti-biotite) solid solutions (same references as before); activity modified end-members for chlorite, amphibole, clinopyroxene and zoisite, matching the measured compositions (i.e. *a*Clc=0.7; *a*Tr=0.7; *a*Di=0.7; *a*Zo=0.75); calcite, kyanite, microcline, muscovite, quartz, titanite, rutile and geikielite pure end-members. A binary solution model for $H_2O$-$CO_2$ fluid (Connolly and Trommsdorff, 1991) was additionally used for the P-T mixed-volatile projection. Following the method illustrated in Groppo et al. (2017), dolomite, margarite, garnet, and wollastonite were excluded from the calculation of the P/T-X($CO_2$) grids and P-T mixed-volatile projection (but not from the calculation of the P/T-X($CO_2$) pseudosections; see above). Corundum was included in the calculation, but the corundum-bearing equilibria are not shown. Calcite and quartz are not considered in excess, thus giving to the modelled P/T-X($CO_2$) grids a more general validity.

### *3.4 Zr-in-Ttn thermometry*

Zr-in-Ttn temperatures were estimated using the calibration of Hayden et al. (2008). Pressure was set to 10.0-10.5 kbar, basing on the results of the pseudosection modelling (see section 5.1 and Table 3); these values are in good agreement with peak-P conditions constrained from the hosting metapelites (Rapa et al., 2016) and more generally at the regional scale for the L-GHS (e.g. Kohn, 2014). We assumed $a$SiO$_2$=1 because quartz is abundant in both the Cal-poor and Cal-rich domains (> 30 vol%); rutile is absent, therefore $TiO_2$ is undersaturated. $a$TiO$_2$=0.60 and $a$TiO$_2$=0.55 were estimated for the Cal-poor and Cal-rich domains respectively, following Ashley and Law (2015) (i.e. the chemical potential of the $TiO_2$ component, $\mu$TiO$_2$, was calculated with Perple_X at the P-T conditions of interest and for the specific bulk composition of each domain, and converted to $a$TiO$_2$ according to the expression proposed by Ashley and Law, 2015). It is worth noting that, although these $a$TiO$_2$ values are lower than that suggested by Corrie and Kohn (2011) and Kohn and Corrie (2011) for similar rutile-absent assemblages (i.e. $a$TiO$_2$=0.85), they give Zr-in-Ttn temperatures that are consistent with the temperatures constrained independently for both the studied sample and the hosting metapelites, using the phase equilibrium approach. Measurement precision for Zr in titanite propagates to temperature uncertainties < 1°C; uncertainties of ± 0.05 in $a$TiO$_2$ and ± 0.5 kbar propagate to additional systematic errors of ± 5 °C and ± 6 °C, respectively.

## 4. Results

### *4.1 Phase diagram modelling*

The titanite-forming, $CO_2$-producing reactions relevant for the studied sample were modelled following the approach described by Groppo et al. (2017), i.e. by combining three different types of phase diagrams, each one useful to investigate different aspects of the $CO_2$-producing history: (i) P/T-X($CO_2$) grids permit identification and full characterization (i.e. stoichiometric balance) of the complete set of $CO_2$-producing univariant and invariant equilibria for the selected compositional system (note that the full characterization of such equilibria is hard to be achieved using pseudosections); (ii) P/T-X($CO_2$) pseudosections allow us to recognise to which of these total equilibria the studied sample is effectively sensible; (iii) mixed-volatile P-T grids (i.e. P-T projections with fluid of variable composition) permit understanding the role of the invariant equilibria, in the case of internally buffered systems. The main novelty introduced here is the inclusion of $TiO_2$ in the system, not just for the calculation of the P/T-X($CO_2$) pseudosections, but also for the calculation of the P/T-X($CO_2$) grids and mixed-volatile P-T projection. In the following, the main results of the P/T-X($CO_2$) pseudosection modelling are described (Fig. 7 and Fig. SM3), while detailed description of the P/T-X($CO_2$) grids and mixed volatile P-T projection is reported in the Supplementary Material and in the Fig. SM2-SM4.

Two P/T-X($CO_2$) pseudosections were calculated in the NCKFMAST-HC system, using the bulk compositions representative of the Cal-poor and Cal-rich domains, respectively (Table 2). The two pseudosections are topologically very similar (Fig. 7 and Fig. SM3). Both the pseudosections are dominated by large tri-variant fields separated by narrow di-variant-fields. Quadri-variant fields appear only at high temperatures or for high values of X($CO_2$). Quartz is stable in the whole P/T-X($CO_2$) range of interest. On the contrary, carbonate stability fields are limited to specific P/T-X($CO_2$) conditions: calcite is stable at X($CO_2$)>0.01 for T=400 °C, and at X($CO_2$)>0.5 for T=720 °C; in the Cal-poor domains its stability field is limited at T<750 °C, whereas in the Cal-rich domains it is predicted to be stable at T>800°C. Dolomite is limited to low-T conditions (T<600°C) or to high $X(CO_2)$ values ($X(CO_2)$>0.6). Plagioclase (with variable composition) is stable over most of the P/T-$X(CO_2)$ range of interest, except for a relatively large field at 0.05<$X(CO_2)$<0.55 and T<730 °C. Scapolite is stable in a wide range of P/T-$X(CO_2)$ conditions at $X(CO_2)$>0.05 and T>470 °C, with few differences between the Cal-rich and Cal-poor domains. Zoisite is limited to $X(CO_2)$<0.55 in both domains.

The chlorite stability field is limited to T<480°C, $X(CO_2)$<0.15; amphibole is confined to an even smaller field at T<450°C, $X(CO_2)$<0.02. Clinopyroxene is predicted to be stable at $X(CO_2)$<0.02 for T=400°C, and at $X(CO_2)$<0.45 for T=700 °C. An almandine-rich garnet is predicted to be stable in both the pseudosections at T>730°C and $X(CO_2)$>0.5; in the Cal-rich domains a grossular-rich garnet is also modelled at T>650 °C, $X(CO_2)$<0.15.

At low T, muscovite is stable over a wide range of $X(CO_2)$ (i.e. $X(CO_2)$=0.03-0.97); its stability field extends up to ~ 630 °C. The breakdown of muscovite coincides with the appearance of K-feldspar. Biotite is predicted to be stable in a wide range of P/T-$X(CO_2)$ conditions, but in the Cal-rich domains its stability field is limited at T<730 °C, whereas in the Cal-poor domains it is predicted to be stable at T>800°C. Finally, the rutile stability field is limited at $X(CO_2)$>0.05 for T=400°C and at $X(CO_2)$>0.9 for T=700°C; titanite is stable at lower $X(CO_2)$ conditions for the same T conditions.

The results of P/T-$X(CO_2)$ grids and pseudosections suggest that the Cal-poor and Cal-rich domains are sensitive to the same univariant and invariant equilibria (Fig. 7 and Fig. SM3):

(i) CAST-HC reaction 5 (Cal + Qz + Rt → Ttn + $CO_2$), that controls the relative stability of rutile vs. titanite;

(ii) CKAS-HC reaction 7 (Cal + Mu + Qz → Kfs + Zo + $H_2O$ + $CO_2$), that marks the appearance of K-

feldspar at X($CO_2$)<0.45 and the breakdown of muscovite;

(iii) NCAS-HC reactions 14a (Cal + Pl + Zo + $CO_2$ → Scp + $H_2O$) and 14b (Scp + Zo + $CO_2$→ Cal + Pl + $H_2O$), that control the relative stability of plagioclase, scapolite, zoisite and calcite;

(iv) CKMAST-HC univariant reactions 85, 90 and 94 and invariant point I35. Reactions 90 (Kfs + Tr + Ttn → Bt + Di + Qz + $H_2O$ + $CO_2$) and 85 (Bt + Cal + Qz → Kfs + Tr + Ttn + $H_2O$ + $CO_2$) limit the stability field of amphibole at T<450°C, whereas reaction 94 (Bt + Cal + Qz → Di + Kfs + Ttn + $H_2O$ + $CO_2$) marks the appearance of clinopyroxene at T > 450°C;

(v) NCKAS-HC univariant reaction 107b and invariant points I44 and I44'. Reaction 107b (Cal + Mu + Qz + Scp → Kfs + Pl + $H_2O$ + $CO_2$) marks the appearance of K-feldspar (and the breakdown of muscovite) at X($CO_2$) > 0.45;

(vi) NCKMAST-HC univariant reactions 133a and 133b and invariant points I65 and I65'. The Cal-absent reactions 133a (Bt + Qz + Scp → Di + Kfs + Pl + Ttn + Zo + $H_2O$ + $CO_2$) and 133b (Bt + Qz + Scp + Zo → Di + Kfs + Pl + Ttn + $H_2O$ + $CO_2$) represent Pl-forming reactions at relatively low values of X($CO_2$).

Among these univariant equilibria, reactions 5, 85, 94 and 133 are titanite-forming and $CO_2$-producing and, except for reaction 5, they involve mineral phases observed in the studied sample (reaction 5 involves rutile as a reactant, which is not observed in the sample). Reactions 85, 94 and 133 are therefore particularly relevant for this study; they emanate from the titanite-bearing invariant points I35, I65 and I65', which correspond to the following "truly" univariant reactions in the P-T mixed-volatile projection (Fig. SM4):

U35: Cal + Kfs + Tr + Ttn → Bt + Di + Qz + F

U65a: Bt + Cal + Pl + Qz + Zo → Di + Kfs + Scp + Ttn + F

U65c: Bt + Cal + Qz + Scp + Zo → Di + Kfs + Pl + Ttn + F,

where F indicates an $H_2O$-$CO_2$ fluid whose composition change as a function of P and T [X($CO_2$)

values are reported in Fig. SM4; e.g. F54 means a $H_2O$-$CO_2$ fluid with $X(CO_2) = 0.54$].

### *4.2 Zr-in-Ttn thermometry and U-Pb data*

Al and Ti in the titanite show an inverse correlation and have a bimodal distribution (Al = 0.23-0.28 a.p.f.u., Ti = 0.70-0.76 a.p.f.u. vs. Al = 0.31-0.35 a.p.f.u., Ti = 0.63-0.69 a.p.f.u.); Zr shows a positive correlation with Al, in agreement with the commonly observed substitution of Al and Zr for Ti (e.g. Franz and Spear, 1985; Kohn, 2017). Overall, the Zr content in the analysed titanite grains is relatively high, spanning between 114 and 237 ppm, corresponding to Zr-in-Ttn temperatures of 729-764 °C (± 15 °C) (Table 3). The frequency distribution of Zr-in-Ttn temperatures shows two main peaks at 731-735°C and 741-750°C, suggesting the existence of two different titanite populations crystallized at T<740°C and at T>740°C, respectively (Fig. 8b). The mean values of Zr-in-Ttn temperatures calculated for the two titanite populations have been compared with the Student *t*-test, which confirmed that they are statistically significant (the results of the statistical analysis are reported in Table SM3).

U ranges between 19 and 207 ppm and shows a linear correlation with Zr, except for two titanite grains which are enriched in U compared to the other grains with similar Zr content. A slight correlation between chemical variations (i.e. Zr content) and ages is observed for most of the analysed titanite. Specifically, titanite grains recording higher temperature of crystallisation are, on average, younger than titanite grains crystallized at lower temperatures (Fig. 8a and Table 3). This feature was also observed by Kohn and Corrie (2011) in similar calc-silicate rocks from the U-GHS, but here it is less pronounced. Overall, the thermometric and geochronological data suggest that the studied calc-silicate rock experienced protracted heating from 730-740°C at 30-26 Ma to peak-T of 750-760 °C at 25-20 Ma (Table 3, Fig. 8 and 9). Two U-enriched titanite grains record younger ages (~15 Ma) and are therefore interpreted as related to a retrograde re-crystallization/re-equilibration event. Zr-temperatures for this retrograde event are poorly constrained, due to the uncertainty in the estimation of pressure; temperatures of ~720°C have been estimated for P = 9 kbar (Table 3).

It is worth noting that the relative distribution of titanite ages is independent from the assumed common Pb composition ($^{207}Pb/^{206}Pb$ = 0.84, following Stacey and Kramers, 1975; however, higher values of common lead such as those used by Kohn and Corrie (2011) for similar calc-silicate rocks (i.e. common Pb $^{207}Pb/^{206}Pb$ = 0.87) would imply slightly older and more spread in absolute ages (Fig. 8c), whereas lower values of common lead would imply younger and more overlapped absolute ages.

## 5. Discussion

### *5.1 P/T-X($CO_2$) prograde evolution*

Most of the observed microstructures reflect either isobaric/isothermal univariant- or invariant-assemblages (i.e. truly-univariant assemblages in the corresponding mixed-volatile P-T projection), thus suggesting that the system remained internally buffered for most of its metamorphic evolution. Moreover, the close similarity between the modal amounts of minerals observed in the Cal-poor and Cal-rich domains and those predicted by the pseudosections (Table 1a), suggests that little or no externally-derived fluid interacted with the rock.

The occurrence of rare amphibole relics preserved in the core of clinopyroxene, tightly constrains the first portion of the P-T-X($CO_2$) evolution in the narrow amphibole stability field, which is limited by the univariant equilibria 85 and 90 (Fig. 7 and Fig. SM3). Moreover, microstructural evidence suggests that clinopyroxene cores grew at the expenses of amphibole + calcite ± quartz (Fig. 3d). This microstructure can be explained by the truly univariant reaction U35 (Cal + Kfs + Tr + Ttn → Bt + Di + Qz + F; corresponding to the isobaric/isothermal invariant point

I35); the same reaction could also explain the rare occurrence of corroded inclusions of K-feldspar in biotite (Fig. 3a).

Once at least one of the reactants of reaction U35 (i.e. amphibole) was completely consumed, the system evolved along the univariant reaction 94 (Bt + Cal + Qz → Di + Kfs + Ttn + $H_2O$ + $CO_2$) until it reached the invariant point I65 (i.e. truly univariant reaction U65a: Bt + Cal + Pl + Qz + Zo → Di + Kfs + Scp + Ttn + F). Reaction U65a could explain the rare occurrence of Na-rich plagioclase preserved as corroded inclusions within scapolite and titanite (Fig. 4e & 5b).

The system departed from the invariant point I65 once one of the reactants was totally consumed. Two possibilities can be discussed: (i) calcite was completely consumed, and the system evolved along the univariant reactions 133a and 133b; (ii) Na-rich plagioclase was completely consumed, and the system evolved along the univariant reaction 94. The first possibility is unlikely, because reactions 133a (Bt + Qz + Scp → Di + Kfs + Pl + Ttn + Zo + $H_2O$ + $CO_2$) and 133b (Bt + Qz + Scp + Zo → Di + Kfs + Pl + Ttn + $H_2O$ + $CO_2$) would have implied the growth of Na-rich plagioclase, which is instead not observed in the studied sample. Rather, reaction 94 (Bt + Cal + Qz → Di + Kfs + Ttn + $H_2O$ + $CO_2$) is more likely to have occurred, and it is consistent with the preservation of biotite relics in clinopyroxene (Fig. 3f). The system then evolved along equilibrium 94 until it reached the truly univariant reaction U65c, corresponding to the isobaric/isothermal invariant point I65'. Reaction U65c (Cal + Bt + Scp + Zo + Qz → Cpx + Kfs + Pl + Ttn + F) was responsible for the growth of most of the Ca-rich plagioclase observed in the sample and is consistent with the occurrence of: (i) inclusions of scapolite, epidote and quartz in Ca-rich plagioclase (Fig. 4a); (ii) inclusions of scapolite, biotite, calcite, quartz and epidote in titanite (Fig. 5a,c,d,e,f); (iii) inclusions of biotite, quartz, calcite and scapolite in the clinopyroxene rims (Fig. 3d,e); (iv) inclusions of scapolite in K-feldspar (Fig. 3b).

Once zoisite was exhausted, the system further evolved along the univariant reaction 94. In Cal-poor domains it is likely that reaction 94 (and its buffering ability) ceased due to the complete consumption of calcite, thus implying that the system entered the Bt + Cpx + Kfs + Pl + Qz + Scp + Ttn stability field that represents the observed peak assemblage. In Cal-rich domains, biotite is probably completely consumed, and the system entered the Cal + Cpx + Kfs + Pl + Qz + Scp + Ttn stability field, in agreement with the observed peak assemblage.

### *5.2 Different episodes of titanite growth and $CO_2$ production*

#### *5.2.1 P-T-t conditions of titanite growth*

It has been demonstrated that, as long as a system remains internally buffered, sudden and volumetrically-significant appearance of new phases and the simultaneous disappearance of previously abundant phases occur at the isobaric/isothermal invariant points, whereas modal changes are only minor along the univariant curves (e.g. Greenwood, 1975; Groppo et al., 2017). In contrast, univariant reactions lead to volumetrically-significant appearance of new phases when the buffering ability of the system ceases due to the complete consumption of one or more reactants.

According to the P-T-X($CO_2$) evolution discussed in section 5.1, both the Cal-poor and Cal-rich domains remained internally buffered up to ~740°C, i.e. up to temperatures slightly higher than I65’. At T > 740°C, the buffering ability of both domains ceased due to the complete consumption of calcite/biotite in the Cal-poor/Cal-rich domains, respectively. It follows that, in both domains, the most important episodes of mineral growth/consumption occurred at the invariant points I35, I65, and I65’, and at T > 740°C once that calcite/biotite were completely consumed in the Cal-poor/Cal-rich domains, respectively. In more detail, the results of thermodynamic modelling predict three main episodes of titanite growth (Fig. 7, Fig. SM3 and Table 1b): (i) a “prograde” titanite-forming event occurred at ~570°C, 7 kbar (i.e. at I65 through reaction U65a), producing a small amount of titanite; (ii) a "near-peak" event occurred at ~730°C, 10 kbar (i.e. at I65' through reaction U65c) and produced ~0.1/0.2 vol% of titanite in the Cal-poor/Cal-rich domains, respectively; (iii) a "peak" titanite-forming event occurred at conditions of 740-765°C, 10.5 kbar through reaction 94, leading to titanite production of ~0.4 vol% in both the Cal-poor and Cal-rich domains.

The prediction of thermodynamic modelling is in very good agreement with the Zr-in-Ttn thermometric results, which suggest the existence of two different titanite populations, crystallized at 730-740 °C ("near-peak" event) and 740-765°C ("peak" event), respectively (Fig. 8b). Moreover, the frequency distribution of the measured Zr-temperatures (i.e. "peak" titanite grains are more frequent than "near-peak" titanites) reflects the volume amounts of the two titanite generations as predicted by the thermodynamic modelling (Table 1b). The lack of titanite grains preserving Zr-temperatures of ~570°C suggests a complete re-equilibration/re-crystallization of the prograde generation of prograde titanite during the "near-peak" and/or "peak" events.

Although the absolute ages of the two titanite generations are partially overlapped, the "peak" titanite grains are, on average, younger than the "near-peak" titanite grains (Fig. 8a,c and Fig. 9), pointing to ages of 30-26 Ma for the "near-peak" event and of 25-20 Ma for the "peak" event, respectively. These two episodes of titanite growth are also documented by the probability density plot (Fig. 8c).

*5.2.2 P-T-t constraints on Himalayan prograde metamorphism compared with previous data*

Our results suggest that the studied calc-silicate rock from the L-GHS experienced a high-T evolution at T > 730°C for at least 10 Ma (from ~30 to ~20 Ma), followed by a retrograde event at ~15 Ma. The good match between the Zr-temperatures obtained from the analysed titanite grains and the P-T conditions independently estimated for the associated metapelites (Rapa et al., 2016), further confirm the validity of our new P-T data. It is to be noted that peak P-T conditions estimated for the associated metapelites suggest incipient partial melting conditions, i.e. most of the associated metapelites still contain abundant white mica, but show microstructural features (e.g. nanogranites included in garnet) indicative of incipient anatexis (Rapa et al., 2016).

Absolute ages obtained for sample 14-53c can be validated against existing monazite chronologic data for the region. Specifically, the prograde evolution of the L-GHS in the Langtang region has been constrained at 35-17 Ma by Kohn et al. (2004, 2005) (early prograde: 37-24 Ma; late prograde: 24-17 Ma), whereas monazite ages younger than ~17 Ma were interpreted as related to final cooling (i.e. melt crystallization). Our titanite ages fit well with these data (Fig. 10), although Kohn et al. (2004, 2005) interpreted both the early prograde and late prograde monazite as sub-solidus phases, thus implicitly assuming that they grew at T < 760°C. However, the same authors (Kohn, 2014) noticed that their interpretation is in contrast with the high Zr-temperatures registered by titanites (Kohn and Corrie, 2011) and discussed the possibility that monazite petrogenesis of core compositions was misinterpreted (e.g., the compositions of monazite cores reflect dissolution-reprecipitation in the presence of melt, not prograde sub-solidus growth).

Compared to the geochronological results obtained by Kohn and Corrie (2011) on titanite from U-GHS calc-silicate rocks from the Annapurna region (some tens of km westward of our studied area), our titanites are few million years younger (peak titanite: ~24 Ma vs. ~20 Ma). This is in agreement with the monazite data discussed by Kohn et al. (2004, 2005) and Corrie and Kohn (2011) (summarized in Kohn, 2014), which show that peak metamorphism in the L-GHS was systematically younger than in the U-GHS.

*5.2.3 Amounts of $CO_2$ produced and implications for the deep carbon cycle*

Both the titanite-forming reactions U65c and 94 are decarbonation reactions and the amount of $CO_2$ produced through these reactions, as predicted by the P/T-X($CO_2$) pseudosections, is of ~0.3 wt% and ~1.5 wt% of $CO_2$ (Fig. 7, Fig. SM3 and Table 1b). These results demonstrate that, during the Himalayan orogeny, $CO_2$ was produced through distinct, short-lived events, which occurred at specific P-T conditions (see also Groppo et al., 2017). In order to extrapolate the amount of produced $CO_2$ to the orogen scale, an estimate of the total volume of this type of $CO_2$-source rocks in the whole Himalayan belt is needed. This is not an easy task, because in most of Himalaya's geological maps calc-silicate rocks are not differentiated from the hosting metapelites. A preliminary estimate of their volume was proposed by Groppo et al. (2017) basing on their own field data from eastern Nepal Himalaya. Here we refine those preliminary estimates by considering additional field data from central-eastern Nepal and literature data from central-western Nepal (Fig. 11). According to these data, calc-silicate rocks similar to those analysed in this study are more abundant in the central and western Nepal Himalaya, where they represent up to ~40 vol% of the whole GHS sequence, than in the eastern Nepal Himalaya, where they represent less than 10 vol% of the whole GHS (~5 vol% on average) (Fig. 11). We therefore consider a conservative (minimum) value of 10 vol% as representative of the total volume of this type of $CO_2$-source rocks with respect to the whole GHS. The entire volume of GHS rocks that experienced Himalayan metamorphism is approximately ~5 x $10^6$ $km^3$ (considering a strike length of 2500 km, an average thickness of 10 km and an average shortening of 200 km; e.g. Long et al., 2011); consequently, the total volume of the $CO_2$-source rocks was ~5 x $10^5$ $km^3$. This estimate would result in a total $CO_2$ production of ~4 x $10^6$ and ~2 x $10^7$ Mt of $CO_2$ during the "near peak" and "peak" events, respectively. Although we have demonstrated that $CO_2$ production occurred in pulses for specific volumes of calc-silicate rocks, extrapolation of the $CO_2$ flux at the orogen scale should integrate these amounts over the time needed for the entire length of GHS to reach the P-T conditions required for $CO_2$ production, which can be approximated to ~20 Ma (e.g. Kohn, 2014).

This would result in metamorphic $CO_2$ flux for the "near-peak" and "peak" events of 0.2 and 1 Mt/yr, respectively. The lack of field evidence for pervasive carbonation and/or massive graphite precipitation in the lithologies structurally overlying the main calc-silicate layers in the GHS (according to our own observations along transects 5-14 in Fig. 11), suggests that most of the $CO_2$ produced through the investigated metamorphic reactions reached Earth's surface without interacting with the hosting rocks.

Sensitivity analysis applied to our estimates (Table 4) shows that the calculated metamorphic $CO_2$ flux is a minimum estimate. Specifically:

(1) $CO_2$ productivity – The studied calc-silicate rock experienced at least two additional $CO_2$-producing events: an "early prograde" one at ~430°C, 4 kbar (i.e. at I35 through reaction U35), which released ~0.8 wt% $CO_2$, and a "prograde" one at ~575°C, 7 kbar (i.e. at I35 through reaction U35), which released ~0.3 wt% $CO_2$ (Fig. 7 and Fig. SM3, SM4). Summed to the amounts produced at "near peak" and "peak" events, a total production of ~3 wt% $CO_2$ would result (Table 1b). Furthermore, compared to other samples of the same calc-silicate type, the studied sample is derived from a marly protolith relatively poor in calcite (Fig. S1). It has been demonstrated that the $CO_2$ productivity of similar calc-silicate rocks derived from Cal-rich marls (see sample 07-22 in Fig. S1) is higher, up to ~5 wt% $CO_2$ (Groppo et al., 2017). On average, a maximum $CO_2$ productivity of ~4 wt% $CO_2$ could be therefore considered for this calc-silicate type.

(2) GHS total volume - Other authors proposed larger estimates for the total volume of GHS rocks (e.g. Kerrick and Caldeira, 1999 considered a value of 9 x $10^5$ $km^3$).

(3) Volume of $CO_2$-source rocks - Our estimate of the volume of this type of $CO_2$-source rocks is a minimum estimate; a value of 25 vol% could be realistically considered as a maximum estimate.

(4) Duration of the $CO_2$-producing process in the whole GHS - The time needed for the entire length of GHS to reach the P-T conditions required for $CO_2$ production is poorly constrained. In the sensitivity analysis, we have considered durations of 20 and 30 Ma as minimum and maximum values, respectively.

(5) Contribution of other $CO_2$-source rocks - Beside the studied calc-silicate rock type, which is the most abundant in the Himalaya and derived from a marly protolith (Fig. SM1), other $CO_2$-source rocks occur in the GHS (see Rolfo et al., 2017 for a review). These $CO_2$-source rocks mostly include calc-silicate –bearing metapelites (i.e. scapolite/anorthite/clinozoisite ± garnet two-micas micaschists and gneisses; Rolfo et al., 2017) deriving from carbonate-rich pelites (Fig. SM1). The $CO_2$-productivity of these lithologies is still unknown, because they have never been investigated in detail; it should be in any case lower than that of the studied calc-silicate rock type, because the protolith contained less calcite. Moreover, their abundance within the GHS is difficult to be constrained, because these rocks are not easy to be recognised in the field, resembling common metapelites (Rolfo et al., 2017). However, according to our experience, their vol% cannot exceed that of the calc-silicate rock type studied in this paper. Overall, the contribution of the other types of $CO_2$-source rocks to the total orogenic $CO_2$ flux should be therefore minor than the $CO_2$ flux estimated for the studied calc-silicate type.

Considering the uncertainties associated with the input values (1) to (4), the metamorphic $CO_2$ flux resulting from the studied calc-silicate type ranges between a minimum value of 1.4 Mt/yr and a maximum value of 9.7 Mt/yr (Table 4). Adding the contribution of the other types of $CO_2$-source rocks (input value 5), the maximum $CO_2$ flux would be 2x higher, resulting in a total metamorphic $CO_2$ flux from the GHS ranging between 1.4 and 19.4 Mt/yr. Further studies on the $CO_2$-source rocks derived from carbonate-rich pelitic protoliths would provide more precise constraints on the $CO_2$ productivity of such lithologies, and consequently on the total orogenic $CO_2$ flux.

Compared to the literature data, our estimated flux of $CO_2$ (1.4 Mt/yr < $CO_2$ flux <19.4 Mt/yr) is slightly lower than the past $CO_2$ flux estimated by Kerrick and Caldeira (1999) (9 Mt/yr < $CO_2$ flux <24 Mt/yr) using the mass loss method. The discrepancy between the two values is probably due to the Kerrick and Caldeira (1999) slight over-estimation of the amount of $CO_2$ released from the $CO_2$-source rocks. Specifically, they assumed that: (i) the whole GHS contributed to the production of $CO_2$ (i.e. $CO_2$-source rocks = 100 vol%), and that (ii) the GHS protolith was a pelite with 5 wt% carbonate (i.e. ~2 wt% $CO_2$)[1] and all carbonate was consumed during metamorphism.

Compared to the present-day $CO_2$ flux estimated on the basis of $CO_2$ degassed from spring waters located along the MCT, our estimated flux of $CO_2$ (1.4 Mt/yr < $CO_2$ flux < 19.4 Mt/yr) is lower than the estimate of Becker et al. (2008) (40 Mt/yr), but similar to that of Evans et al. (2008) (8.8 Mt/yr). Becker et al. (2008) based their estimates on data from the Marsyandi basin of central Nepal, which cover about ~4800 $km^2$, whereas the data used by Evans et al. (2008) refer to the larger Narayani basin, covering ~40000 $km^2$ and comprising the Marsyandi basin itself. According to the Evans et al. (2008) data, the contribution of the Marsyandi spring system to the total flux of $CO_2$ is nearly one order of magnitude higher than the contributions of the other spring systems of the Narayani basin. It therefore appears that the Becker et al. (2008) extrapolation to the whole orogen is biased by the exceptional high $CO_2$ flux from the Marsyandi basin, whereas the estimate by Evans et al. (2008) probably approximates better the contribution of the majority of the spring systems of the Himalaya. The similarity between our estimated flux of $CO_2$ and that actually measured from spring waters thus suggests that $CO_2$-producing processes similar to those described in the present work still occur along the active Himalayan orogen (e.g. Girault et al., 2014).

[1] Note that although in the Kerrick and Caldeira's paper a pelite with 5 wt% $CO_2$ (instead of 5 wt% carbonate) was mentioned as GHS protolith (which would correspond to 12.5 wt% carbonate), the authors explicitly referred to the Bickle's paper (Bickle, 1996) in which a pelite with 5 wt% carbonate was considered.

To our knowledge, this is the first time that $CO_2$-producing metamorphic processes occurring during an orogenic event have been fully characterized in term of P-T conditions, time, duration and amounts of $CO_2$ produced. Although the volumes of the $CO_2$-source rocks involved in such processes could be potentially better constrained in the future, we suggest that these metamorphic $CO_2$ fluxes should be considered in any future attempts of estimating the global budget of non-volcanic carbon fluxes from the lithosphere (e.g. Morner and Etiope, 2002).

**ACKNOWLEDGMENTS**

Carl Spandler (James Cook University, Australia) and Allen Kennedy (Curtin University, Australia) are kindly acknowledged for providing the MKED1 and OLT1 titanite samples. A. Skelton, A. Galy and an anonymous reviewer provided constructive comments on an earlier version of this paper. M. Kohn and A. Skelton are gratefully acknowledged for their careful and constructive reviews which significantly improve the manuscript. The instrument used to obtain the SEM-EDS data was acquired thanks to a grant from the Compagnia di San Paolo, Torino. The study was supported by the University of Torino (Ricerca Locale, ex-60% - 2014, 2015 funds) and Compagnia di San Paolo (University of Torino, Call 1, Junior PI Grant: TO_Call1_2012_0068), by the Italian Ministry of University and Research (PRIN 2011: 2010PMKZX7) and by Ev-K2-CNR (SHARE Project). The financial support of European Research Council for the Consolidator Grant ERC-2013-383 CoG Proposal No. 612776 — CHRONOS (DP) is also acknowledged.

## Figure captions

**Fig. 1 - (a)** Geological sketch map of the central–eastern Nepal Himalaya, showing major tectono-metamorphic units (modified from Goscombe and Hand, 2000; He et al., 2015; Wang et al., 2016 and based on our own data). The white rectangle indicates the location of (b). 1: Siwalik deposits; 2: Lesser Himalayan Sequence; 3: Lower Greater Himalayan Sequence; 4: Upper Greater Himalayan Sequence; 5: Tethyan Sedimentary Sequence. MFT, Main Frontal Thrust; MBT, Main Boundary Thrust; MCT: Main Central Thrust; STDS: South Tibetan Detachment System. The inset locates the study area in the framework of the Himalayan chain. **(b)** Geological map of the Langtang and Gosainkund–Helambu regions in central Nepal Himalaya (modified from Rapa et al., 2016) and location of the studied sample 14-53c (yellow star). The white stars indicate the location of L-GHS samples dated by Kohn et al. (2004, 2005) (monazite U-Pb ages), and discussed in the text. GSZ, Galchi Shear Zone (from He et al., 2015); MCT, Main Central Thrust; LT, Langtang Thrust. **(c)** Outcrop images of the ~2 m -thick calc-silicate level from which the studied sample 14-53c was collected. The calc-silicate level shows a banded structure, being characterized by dm-thick layers with different mineral assemblages: (a) quartz + clinopyroxene + calcic plagioclase; (b) garnet + calcic plagioclase + clinopyroxene + epidote; (c) quartz + clinopyroxene + calcic plagioclase + K-feldspar + scapolite ± calcite; (d) quartz + garnet + scapolite + clinopyroxene + calcic plagioclase.

**Fig. 2** – Processed micro-XRF map of the whole thin section 14-53c, with distinction of Cal-rich and Cal-poor domains. Note that the fine-grained nature of epidote hampered its discrimination in the micro-XRF map. Calcite-poor domains consist of (vol%): quartz (45%), K-feldspar (15%), clinopyroxene (11%), plagioclase (10%), scapolite (9%), zoisite (4%), biotite (4%), titanite (2%), minor calcite (<<1%) and accessory tourmaline. Calcite-rich domains are characterized by higher modal amounts of calcite, scapolite and plagioclase and lower amounts of K-feldspar, quartz and biotite (quartz: 34%; scapolite: 17%; plagioclase: 14%; clinopyroxene: 14%; K-feldspar: 9%; zoisite: 6%; calcite: 1%; biotite: 3%; titanite: 2%).

**Fig. 3** – Representative microstructures involving K-feldspar and clinopyroxene. **(a)** K-feldspar and quartz inclusions in a biotite flake (Back Scattered Electron image, BSE). **(b)** K-feldspar includes rounded quartz and scapolite crystals (Crossed Polarized Light: XPL). **(c)** Rounded inclusion of clinopyroxene in K-feldspar (XPL). **(d)** Clinopyroxene granoblast in a Cal-rich domain (XPL). The inset shows a detail of the amphibole + calcite + quartz polymineralic inclusion hosted in the clinopyroxene core, with corroded and lobated margins against clinopyroxene (BSE). **(e)** Clinopyroxene granoblast in a Cal-poor domain, with scapolite, titanite and quartz inclusions in the rims. K-feldspar in the matrix exhibits straight contacts with it (BSE). **(f)** Rounded biotite included in the clinopyroxene core in a Cal-rich domain (Plane Polarized Light: PPL).

**Fig. 4** – Representative microstructures involving plagioclase and scapolite. **(a)** Scapolite and quartz inclusions in plagioclase, showing lobated and rounded margins against it (XPL). **(b)** Plagioclase granoblast in a Cal-rich domain, including clinopyroxene, K-feldspar and quartz (XPL). **(c)** Plagioclase in a Cal-poor domain, growing at the expenses of clinopyroxene (XPL). **(d)** Plagioclase from a Cal-rich domain includes rounded quartz and calcite. It is replaced by an aggregate of epidote + calcite at its margins (BSE). **(e)** Relict of plagioclase (bytownite) replaced by scapolite (XPL). The inset highlights the corroded margins of plagioclase (BSE). **(f)** Calcite growing on scapolite in a Cal-rich domain. (XPL). **(g)** Epidote + quartz + calcite symplectitic aggregate growing at the expense of scapolite (retrograde microstructure) (XPL). **(h)** Detail of (g) (BSE). **(i)** Plagioclase including K-feldspar and quartz, replaced by an epidote rim (retrograde microstructure) in a Cal-poor domain (XPL).

**Fig. 5** – Representative microstructural relationships between titanite and the other rock-forming minerals (BSE). a,e,f: Cal-poor domain; b,c,d: Cal-rich domain.

**Fig. 6** – Compositional diagrams for scapolite, plagioclase, clinopyroxene and amphibole.

**Fig. 7** – **(a)** P/T-X($CO_2$) pseudosection calculated for the Cal-poor domain, in the system NCKMFAST-HC along a P/T gradient reflecting the P-T path followed by the hosting metapelites (Rapa et al., 2016; P (bar) = 20.5 T(K) – 10294). White, light- and dark-grey fields are di-, tri- and quadri-variant fields, respectively; the narrow divariant fields correspond to the univariant reactions in the corresponding P/T-X($CO_2$) sections (see Fig. SM2). Dashed fields are dolomite-, garnet-, and/or chlorite-bearing fields, not relevant for this study (because these phases are not observed in the studied sample). The purple arrow approximates the P/T-X($CO_2$) internally buffered fluid evolution as constrained by the relevant microstructures (see Section 5.1). The purple box constrains the Zr-in-Ttn temperatures obtained from the analysed titanite grains. **(b)** Same isobaric P/T–X($CO_2$) pseudosection as in (a): the univariant and invariant equilibria relevant to the Cal-poor domain are outlined (reactions labels and colors as in Fig. SM2). Note that the univariant curves and invariant points overlap the (narrow) divariant fields and (short) univariant lines of the pseudosection. The inset shows the variation in titanite (vol%) and $CO_2$ (wt%) amounts along the inferred P/T-X($CO_2$) path (purple arrow in (a)) and highlights the main prograde episodes of titanite growth and $CO_2$ production (see Section 5.2).

**Fig. 8** – **(a)** Temperature-time diagram for sample 14-53c, showing gradually increasing temperatures between 28 and 19 Ma and a retrograde cooling at ~15 Ma (grey symbols refer to the retrograde titanite generation). Errors on temperatures are ± 10°C (see Section 3.4 for Discussion). **(b)** Frequency distribution of Zr-in-Ttn temperatures; the two titanite grains interpreted as retrograde are not considered, due to the uncertainties in the estimate of their temperature of crystallization. **(c)** Probability density plot for U-Pb ages (common Pb $^{207}Pb/^{206}Pb$ = 0.84); although partially overlapped, the "near peak" and "peak" titanite generations are clearly

distinguishable. The dotted line refers to titanite ages obtained for a higher common Pb value ($^{207}Pb/^{206}Pb$ = 0.9).

**Fig. 9 –** Inverse isochron plot of the titanite U–Pb data, plotted using ISOPLOT (Ludwig, 2001). The numbers marked along the concordia line are ages in Ma. Color coded according to Zr-in-Ttn temperatures (as in Fig. 8).

**Fig. 10** – Comparison between monazite and titanite ages from the L-GHS unit in the Langtang-Gosainkund region of central Nepal (see Fig. 1b for the location of the analysed samples). **(a)** Probability distribution of monazite ages from Kohn et al. (2004, 2005). **(b)** Probability distribution of titanite ages (this study). Note the good match between the two sets of data, which reinforces the validity of our results. Color coded as in Fig. 8.

**Fig. 11** – (a) Simplified tectonic map of the Himalayan-Tibetan orogen, with location of the cross-sections reported in (b). Modified from He et al. (2015) and Wang et al. (2016). (b) Fourteen schematic profiles across the Himalayan metamorphic core in Nepal, representing the main tectono-metamorphic units, major structures, and abundance of Cpx + Kfs + Scp + Pl ± Qz ± Cal calc-silicate rocks (vol%, below each profile). The profiles are constructed from detailed mapping along each sections by our group (cross-sections 5 to 14; see also Goscombe et al., 2006 for cross-sections 10, 12, 13, 14, and Leloup et al., 2015 for cross-section 7) and by other authors: 1: Yakymchuk and Godin, 2012; 2: Iaccarino et al., 2015; 3: Martin et al., 2010; Corrie and Kohn, 2011; 4: Colchen et al., 1986.

Fig. 1

Click here to download high resolution image

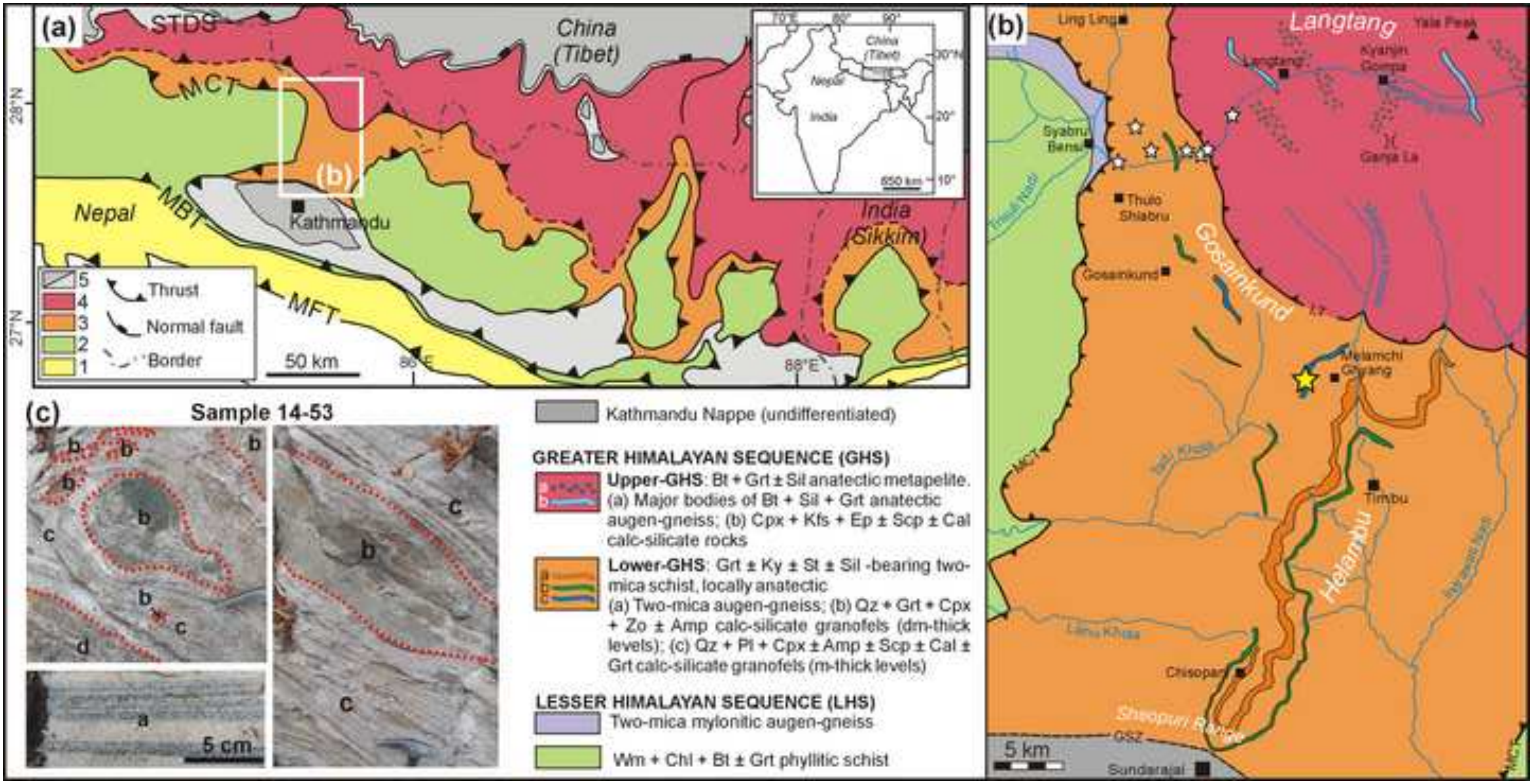

Fig. 2
Click here to download high resolution image

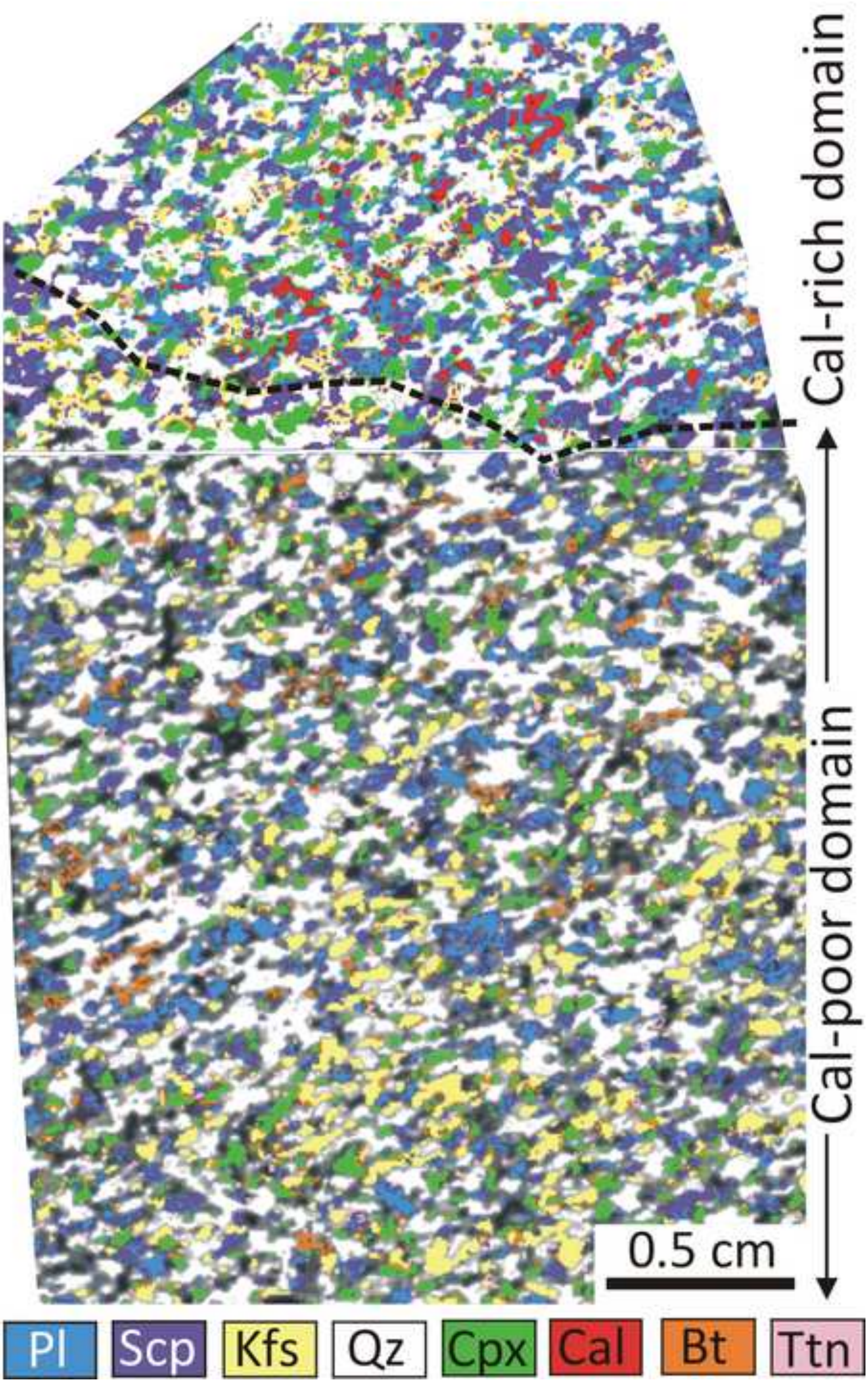

Fig. 3

Click here to download high resolution image

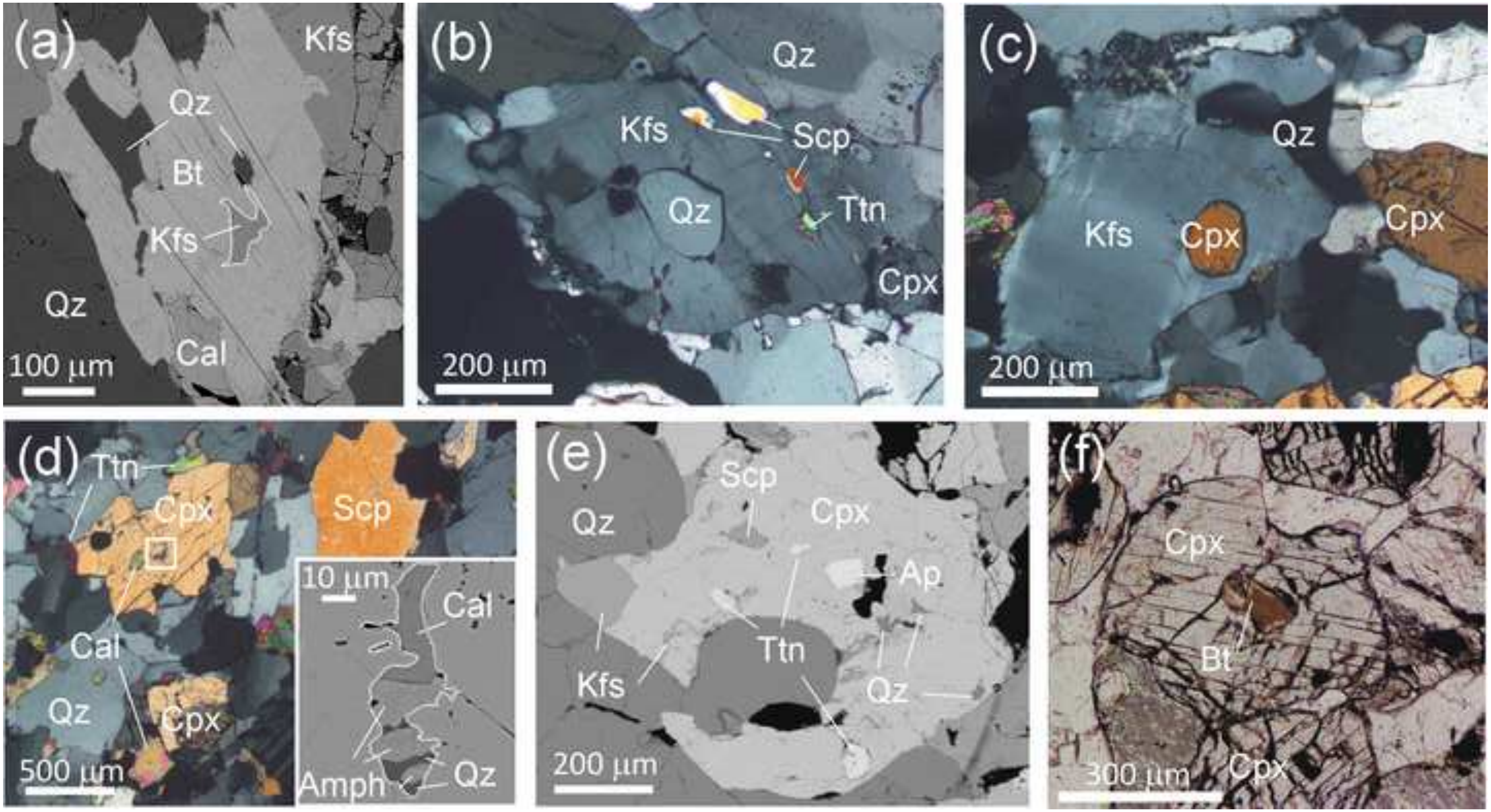

Fig. 4

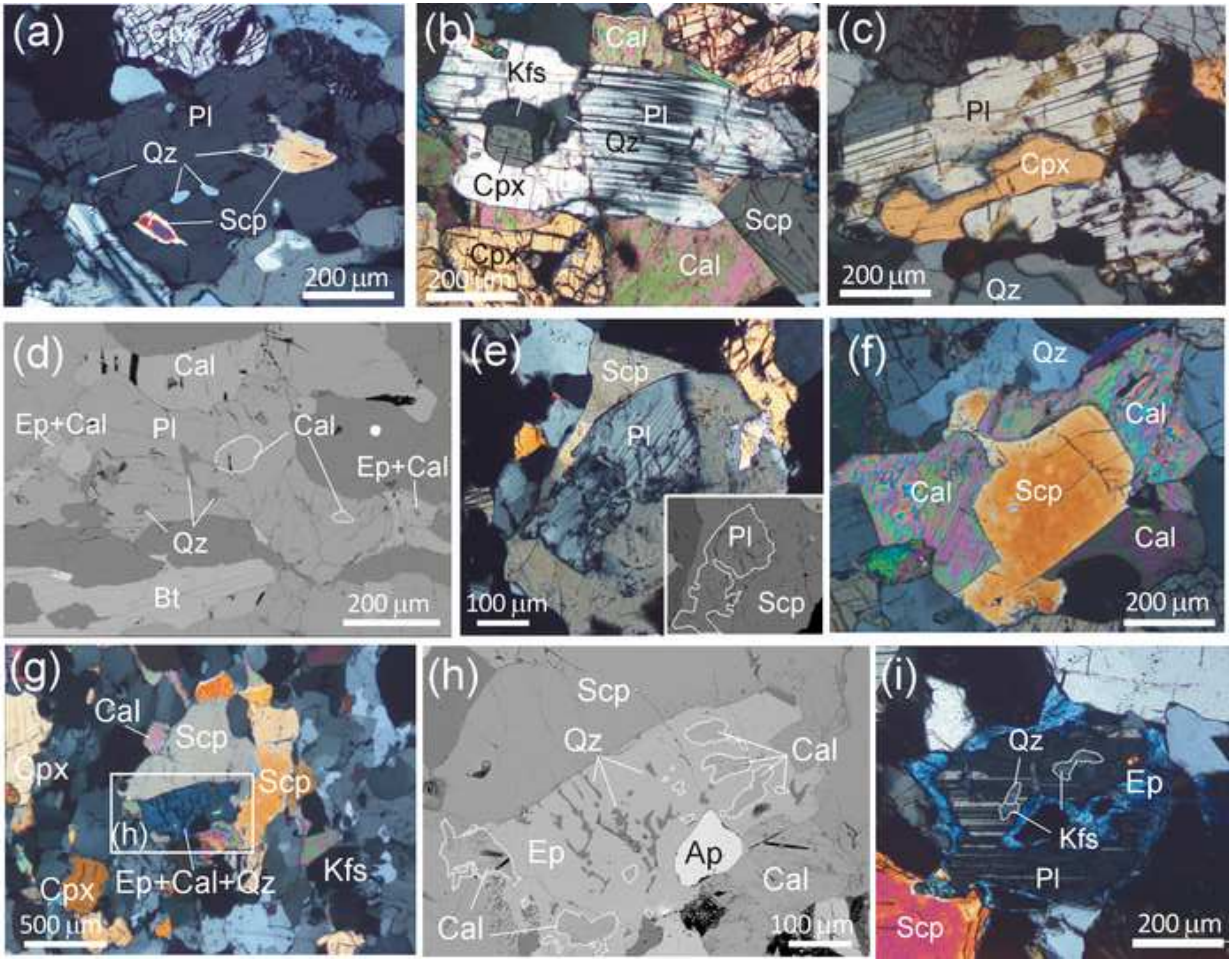

Fig. 5

Click here to download high resolution image

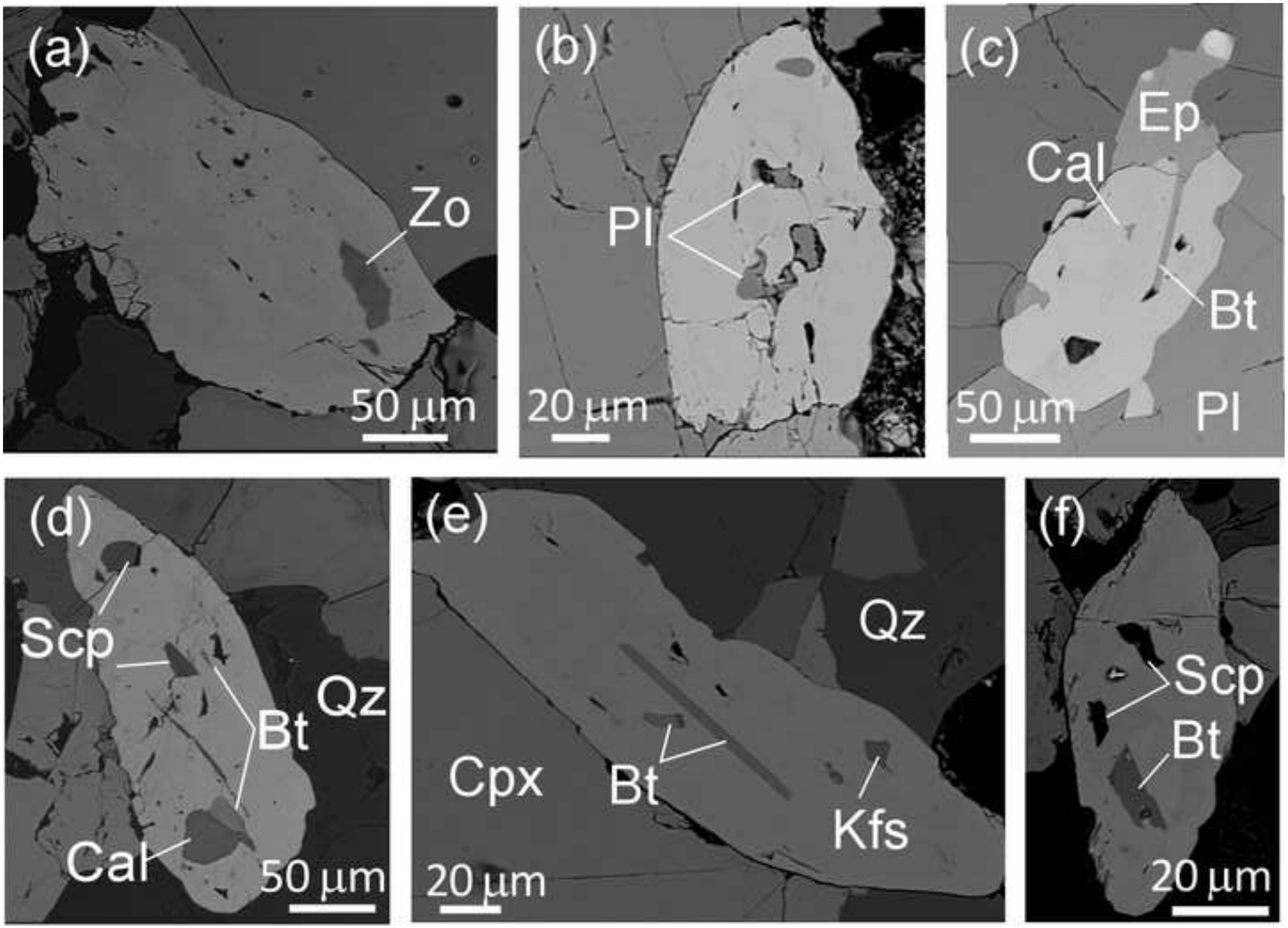

Fig. 6
Click here to download high resolution image

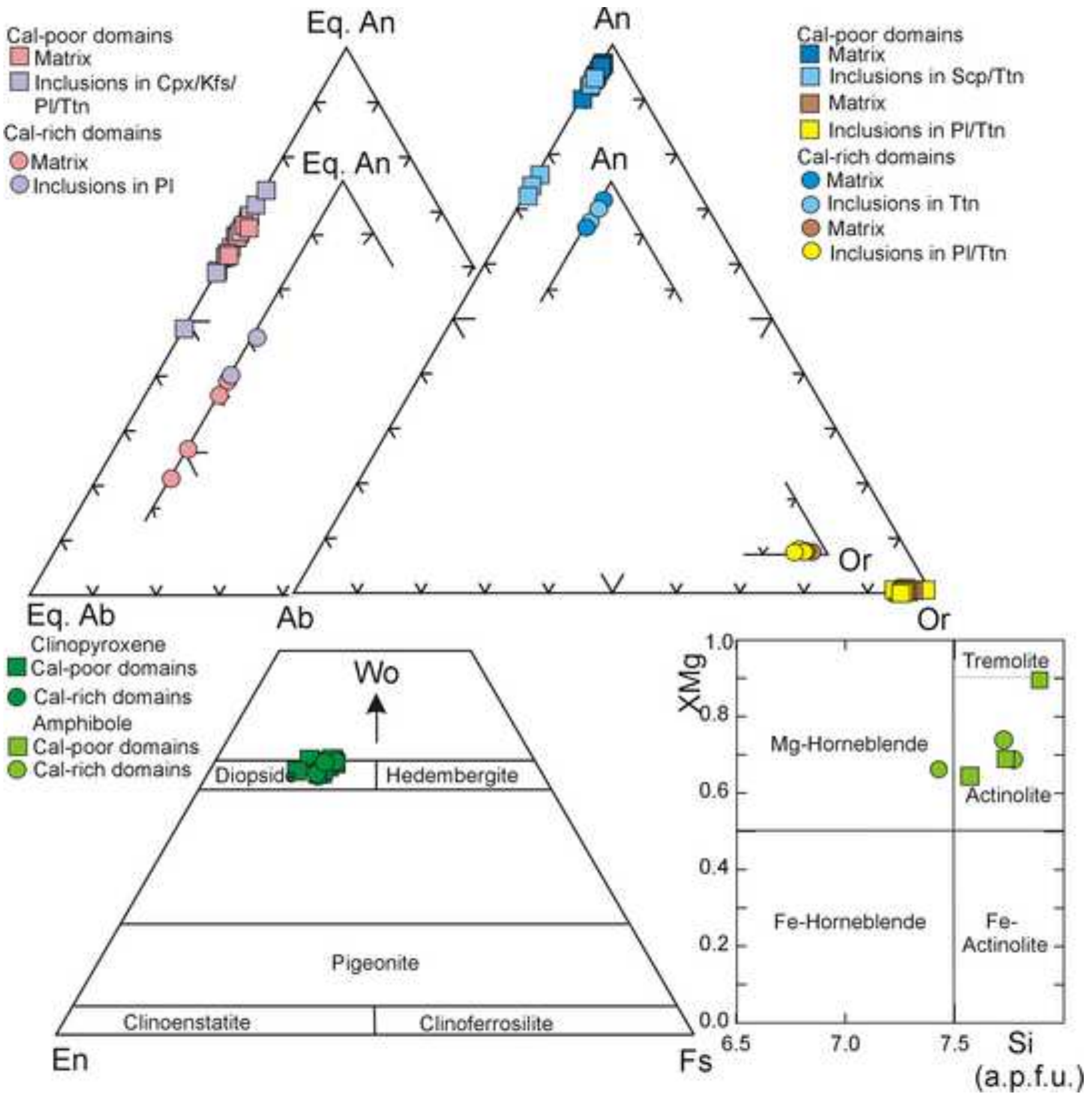

**Fig. 7**
Click here to download high resolution image

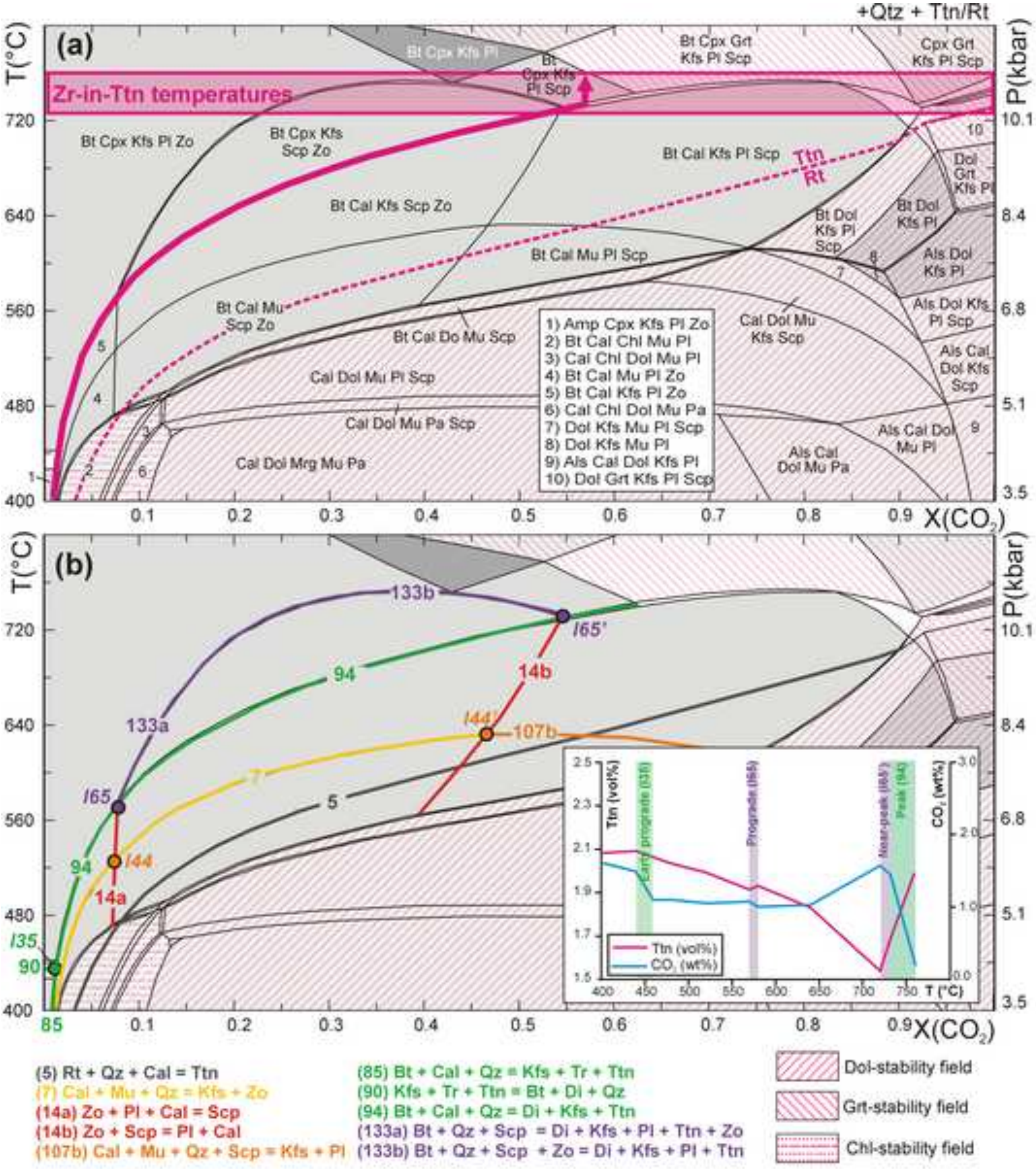

Fig. 8

Click here to download high resolution image

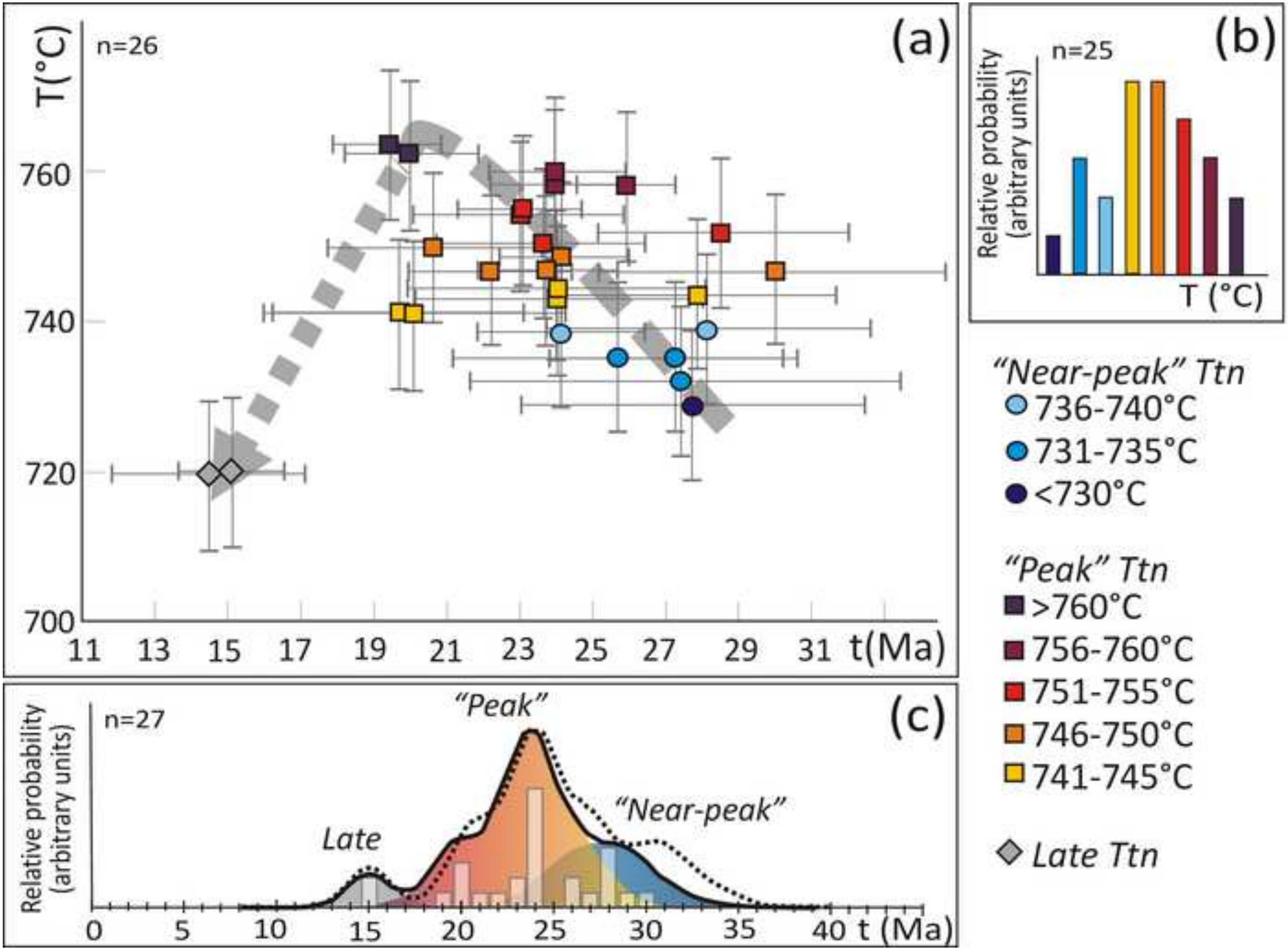

Fig. 9

Click here to download high resolution image

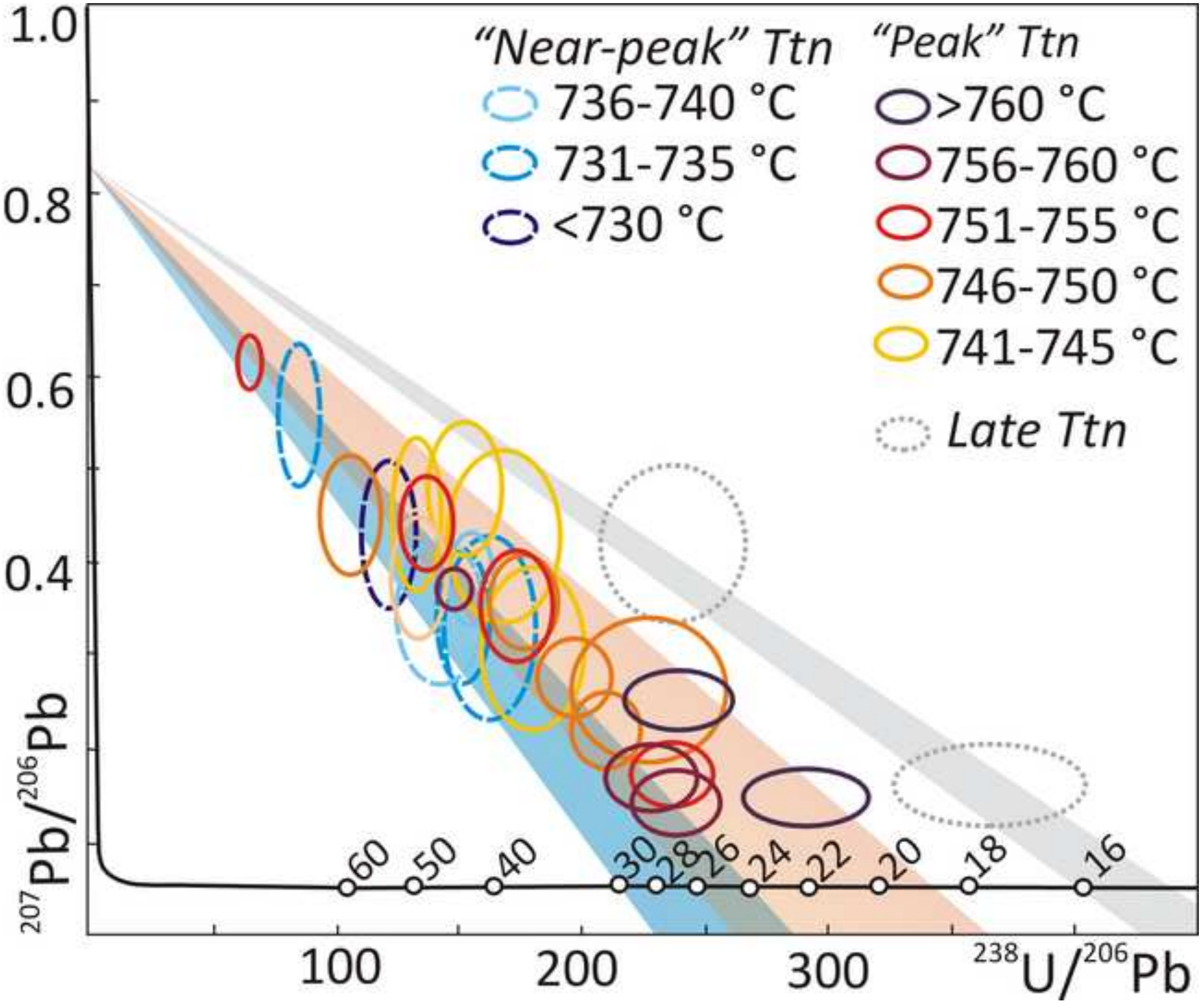

Fig. 10

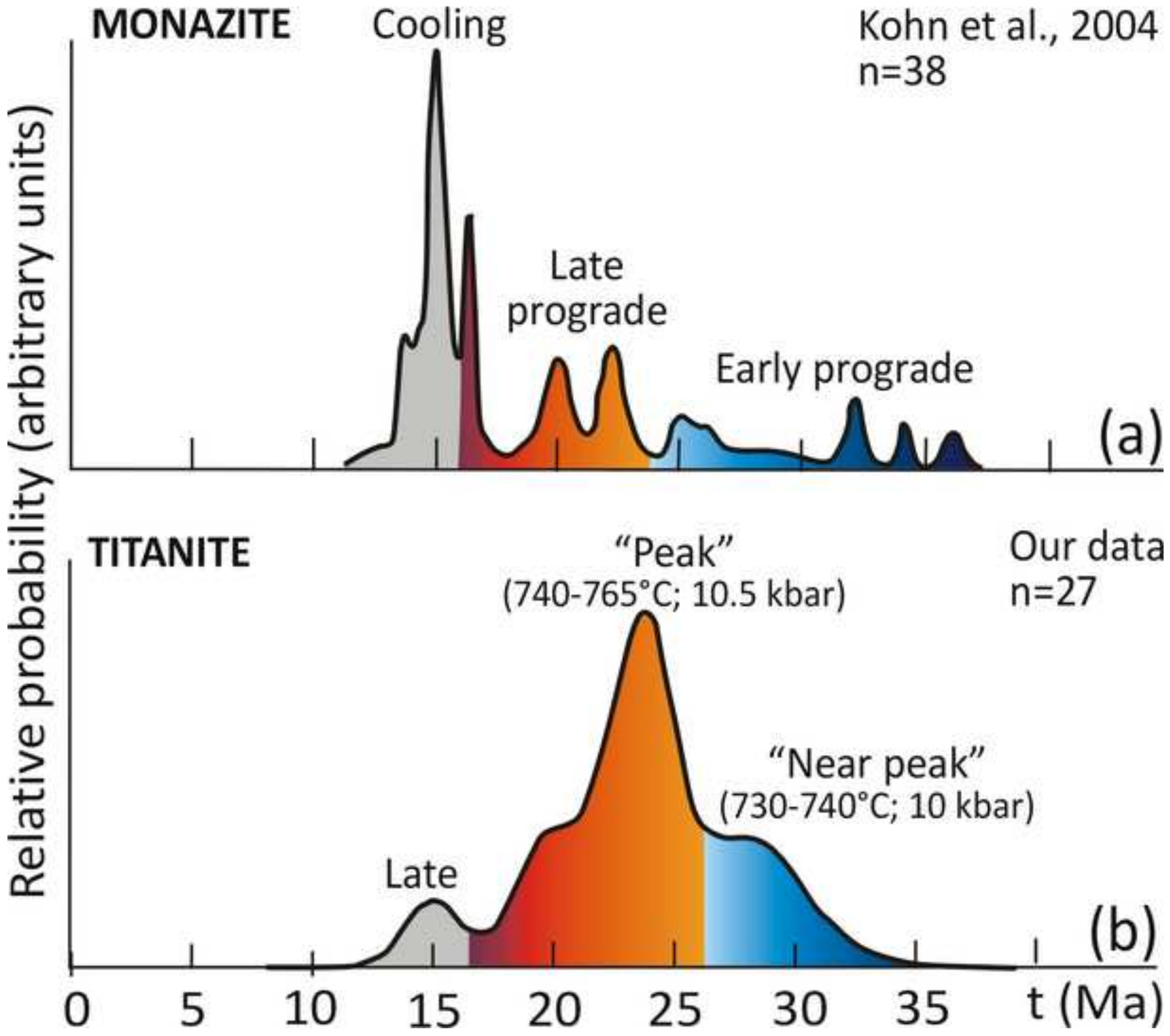

Fig. 11
Click here to download high resolution image

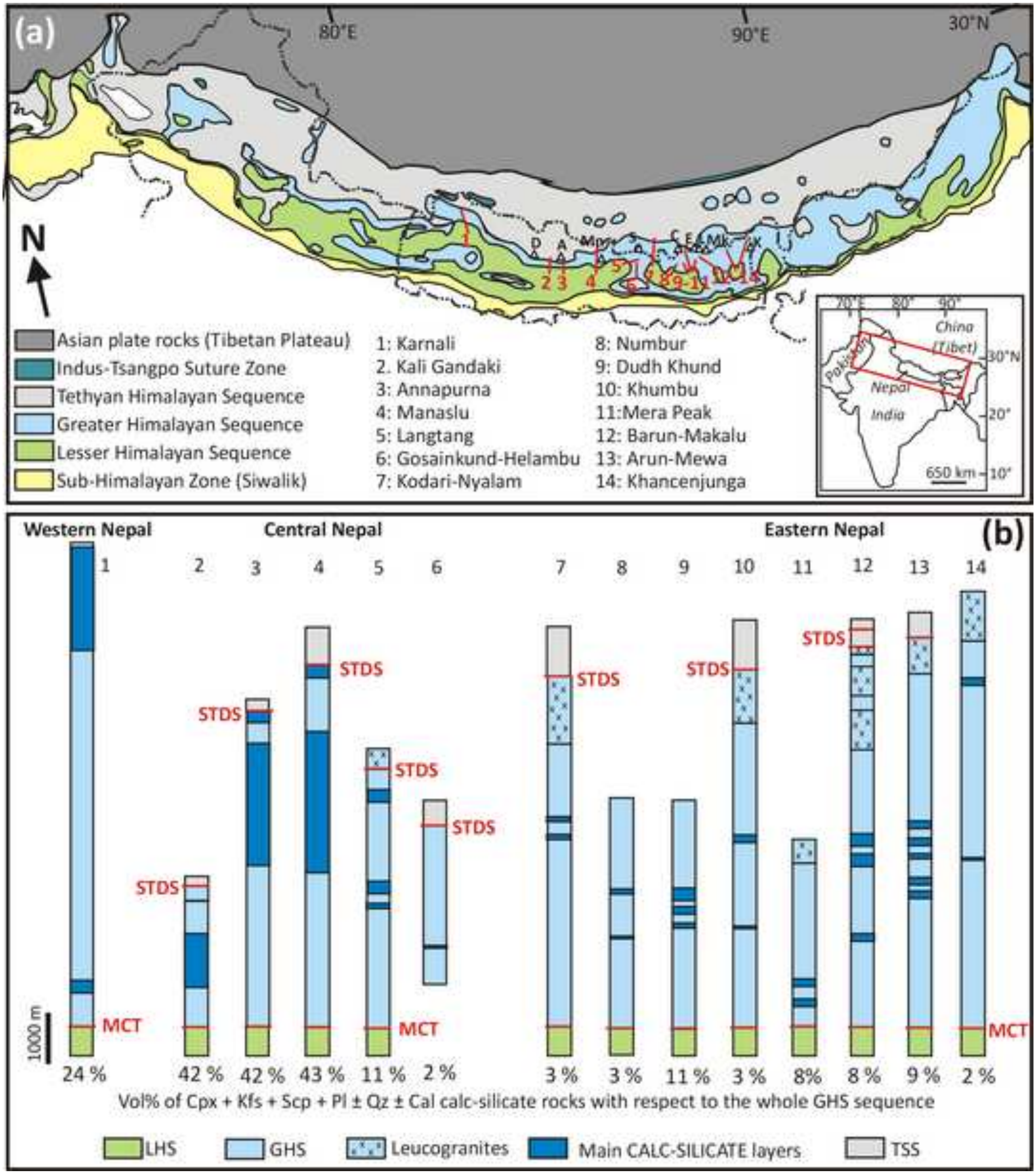

**Table 1**
**Click here to download Table: Table 1.xlsx**

**Table 1.**
**Mineral modes and predicted amounts of titanite and $CO_2$**

(a) Predicted vs. observed modal amount (vol%) of minerals

| | Cal-poor domain | | Cal-rich domain | |
|---|---|---|---|---|
| | Predicted at 750°C, X($CO_2$)=0.57 | Observed | Predicted at 750°C, X($CO_2$)=0.57 | Observed |
| Bt | 3.0 | 4 | - | 3 |
| Cal | - | - | 1.8 | 1 |
| Cpx | 12.8 | 11 | 17.0 | 14 |
| Kfs | 16.1 | 15 | 10.9 | 9 |
| Pl | 19.1 | 10 | 22.7 | 14 |
| Qz | 43.5 | 45 | 31.5 | 34 |
| Scp | 3.5 | 9 | 14.6 | 17 |
| Ttn | 2.0 | 2 | 1.5 | 2 |
| Zo | - | 4 | - | 6 |

(b) Predicted amounts of titanite and $CO_2$ produced during different events

| | Cal-poor domain | | Cal-rich domain | |
|---|---|---|---|---|
| | Ttn (vol%) | $CO_2$ (wt%) | Ttn (vol%) | $CO_2$ (wt%) |
| early prograde | -0.1 | 0.50 | -0.1 | 0.85 |
| prograde | 0.02 | 0.10 | 0.05 | 0.30 |
| near-peak | 0.1 | 0.15 | 0.2 | 0.25 |
| peak | 0.4 | 1.25 | 0.4 | 1.45 |
| Total | 0.4 | 2.0 | 0.6 | 2.9 |

**Table 2**

**Click here to download Table: Table 2.xlsx**

**Table 2.**

**Bulk compositions (mol%)**

| | Cal-poor domain | Cal-rich domain |
|---|---|---|
| $SiO_2$ | 74.81 | 67.91 |
| $TiO_2$ | 0.92 | 0.62 |
| $Al_2O_3$ | 6.76 | 8.65 |
| MgO | 3.56 | 3.74 |
| FeO | 1.93 | 2.00 |
| CaO | 9.77 | 15.29 |
| $Na_2O$ | 0.32 | 0.65 |
| $K_2O$ | 1.93 | 1.15 |
| Total | 100.00 | 100.00 |

**Table 3**
**Click here to download Table: Table 3.xlsx**

**Table 3.**
**Zirconium temperature, isotopic data and calculated ages (± 2σ) for sample 14-53c**

| Analysis | *a* ($TiO_2$) | P (GPa) | Zr (ppm) | U (ppm) | T (°C)[§] | $^{238}U/^{206}Pb$ | ± 2σ | $^{207}Pb/^{206}Pb$ | ± 2σ |
|---|---|---|---|---|---|---|---|---|---|
| *"Near peak" Ttn generation* | | | | | | | | | |
| A.1 ttn2 | 0.55 | 1 | 133 | 19 | 729 | 120.34 | 8.69 | 0.428 | 0.064 |
| 4.g ttn16 | 0.6 | 1 | 130 | 32 | 732 | 85.25 | 6.54 | 0.551 | 0.061 |
| B.2 ttn | 0.6 | 1 | 139 | 34 | 735 | 151.52 | 9.41 | 0.333 | 0.058 |
| 8.d ttn28 | 0.55 | 1 | 140 | 42 | 735 | 162.34 | 15.81 | 0.325 | 0.080 |
| N.1 ttn | 0.6 | 1 | 145 | 28 | 737 | 154.80 | 8.39 | 0.378 | 0.040 |
| L.1 ttn2 | 0.6 | 1 | 149 | 29 | 739 | 143.06 | 14.74 | 0.342 | 0.065 |
| *"Peak" Ttn generation* | | | | | | | | | |
| 8.m ttn34 | 0.6 | 1.05 | 140 | 31 | 741 | 168.63 | 19.05 | 0.423 | 0.078 |
| G.1 ttn1 | 0.6 | 1.05 | 142 | 37 | 742 | 152.91 | 12.63 | 0.471 | 0.061 |
| D.1 ttn1 | 0.6 | 1.05 | 145 | 28 | 743 | 132.63 | 9.15 | 0.447 | 0.065 |
| 3.c ttn10 | 0.6 | 1 | 164 | 67 | 744 | 133.33 | 10.13 | 0.383 | 0.054 |
| 8.c ttn27 | 0.6 | 1 | 169 | 73 | 745 | 180.18 | 17.53 | 0.305 | 0.073 |
| H.2 ttn | 0.6 | 1.05 | 157 | 41 | 747 | 177.62 | 11.99 | 0.354 | 0.042 |
| N.2 ttn 1 | 0.6 | 1.05 | 157 | 57 | 747 | 196.08 | 12.30 | 0.270 | 0.033 |
| 6.a ttn21 | 0.6 | 1 | 177 | 62 | 747 | 106.04 | 10.57 | 0.444 | 0.053 |
| 7.a ttn24 | 0.6 | 1 | 183 | 116 | 749 | 209.64 | 11.87 | 0.213 | 0.033 |
| 3.a ttn9 | 0.6 | 1.05 | 167 | 57 | 750 | 227.27 | 25.83 | 0.259 | 0.063 |
| H.1 ttn2 | 0.6 | 1.05 | 169 | 63 | 751 | 136.43 | 8.38 | 0.441 | 0.043 |
| F.1 ttn1 | 0.6 | 1 | 193 | 96 | 752 | 65.45 | 3.94 | 0.610 | 0.024 |
| I.1 ttn2 | 0.6 | 1.05 | 181 | 82 | 754 | 173.91 | 12.40 | 0.349 | 0.050 |
| 4.a ttn12 | 0.6 | 1.05 | 183 | 89 | 755 | 236.97 | 14.04 | 0.167 | 0.030 |
| H.1 ttn1 | 0.6 | 1.05 | 195 | 112 | 758 | 238.66 | 14.81 | 0.135 | 0.028 |
| A.1 ttn | 0.55 | 1 | 237 | 207 | 758 | 147.93 | 5.47 | 0.367 | 0.018 |
| L.1 ttn1 | 0.6 | 1.05 | 201 | 104 | 760 | 227.79 | 15.05 | 0.165 | 0.031 |
| E.2 ttn | 0.6 | 1.05 | 211 | 130 | 762 | 239.81 | 17.25 | 0.247 | 0.028 |
| G.1 ttn3 | 0.6 | 1.05 | 216 | 140 | 764 | 291.55 | 21.25 | 0.142 | 0.024 |
| *Late Ttn generation* | | | | | | | | | |
| C.1.l ttn3 | 0.6 | 0.9 | 130 | 73 | 720 | 238.10 | 23.81 | 0.414 | 0.069 |
| I.1 ttn1 | 0.6 | 0.9 | 133 | 91 | 721 | 366.30 | 32.20 | 0.155 | 0.037 |

[§]Uncertanties in T is assumed to be ±10°C.

Table 4
Click here to download Table: Table 4.xlsx

**Table 4.**
**Sensitivity analysis for the metamorphic $CO_2$ flux from the Greater Himalayan Sec**

| Variables | Values used in this paper | Max duration of $CO_2$-producing process |
|---|---|---|
| V GHS ($km^3$) | 5.0E+06 | 5.0E+06 |
| Vol% calc-silicate rocks (%) | 10 | 10 |
| $CO_2$ productivity (Wt% $CO_2$) | 3 | 3 |
| Duration of the $CO_2$-producing process (Ma) | 20 | 30 |
| **$CO_2$ flux* (Mt/yr)** | **2.0** | **1.4** |
| **Total $CO_2$ flux[§] (Mt/yr)** | **< 4** | **< 3** |

* $CO_2$ flux from the studied calc-silicate rock type (i.e. derived from marly protoliths); [§] Total $CO_2$ flux
rich pelitic protoliths). Grey cells correspond to the variable parameters.